\documentclass[twocolumn,reprint,amsmath,amssymb,aps]{revtex4-2}
\usepackage{graphicx,dcolumn,bm,float,soul,xcolor,physics}

\newcommand{\Hh}{\hat{H}}
\newcommand{\ah}{\hat{a}}
\newcommand{\Sh}{\hat{S}}
\newcommand{\roh}{\hat{\rho}}
\newcommand{\Nh}{\hat{N}}
\newcommand{\bh}{\hat{b}}
\newcommand{\Ah}{\hat{A}}
\newcommand{\Bh}{\hat{B}}
\newcommand{\Xh}{\hat{X}}

\newcommand{\expv}[1]{\langle #1 \rangle}
\newcommand{\om}{\omega_0}

\begin{document}

\preprint{APS/123-QED}

\title{Dynamics of a Generalized Dicke Model for Spin-1 Atoms}

\author{Ofri Adiv}
\author{Scott Parkins}
\affiliation{Department of Physics, University of Auckland, Auckland 1010, New Zealand}
\affiliation{Dodd-Walls Centre for Photonic and Quantum Technologies, New Zealand}

\date{\today}

%: --------------------------------------------------------------
%:                  Abstract
% --------------------------------------------------------------
\begin{abstract}
The Dicke model is a staple of theoretical cavity Quantum Electrodynamics (cavity QED), describing the interaction between an ensemble of atoms and a single radiation mode of an optical cavity. It has been studied both quantum mechanically and semiclassically for two-level atoms, and demonstrates a rich variety of dynamics such as phase transitions, phase multistability, and chaos. In this work we explore an open, spin-1 Dicke model with independent co- and counter-rotating coupling terms as well as a quadratic Zeeman shift enabling control over the atomic energy-level structure. We investigate the stability of operator and moment equations under two approximations and show the system undergoes phase transitions. To compliment these results, we relax the aforementioned approximations and investigate the system semiclassically. We show evidence of phase transitions to steady-state and oscillatory superradiance in this semiclassical model, as well as the emergence of chaotic dynamics. The varied and complex behaviours admitted by the model highlights the need to more rigorously map its dynamics.
\end{abstract}

\maketitle

%: --------------------------------------------------------------
%:                  Intro
% --------------------------------------------------------------
\section{Introduction}\label{sec:intro}

Since the advent of quantum theory, the study of many-body atomic systems and their interactions with light have provided fundamental insights into a variety of complex behaviors. Among these is Dicke superradiance \cite{dicke}, where a cloud of initially excited atoms radiates collectively with high intensity in a pulse of short duration, provided inter-atomic separations within the cloud are much smaller than the wavelength of the emitted light. This is in stark contrast to spontaneous emission by a cloud of independent, distantly separated atoms, the intensity of which decays exponentially over a much longer timescale set by the atomic excited state lifetime.

A useful experimental and theoretical framework to explore such systems is cavity QED, since it allows for fine control over the relevant modes of the electromagnetic field and system parameters. Hepp and Lieb \cite{hepplieb} studied superradiance by considering an ensemble of $N$ two-level emitters confined to an optical cavity and interacting with only one of its modes, in what is referred to as the ``Dicke model.'' Like the aforementioned variation between spontaneous emission and Dicke superradiance, this first version of the Dicke model was shown to undergo a quantum phase transition dependent on the atom-cavity coupling strength. With increasing coupling strength the model transitioned from the Normal Phase (NP), with an unexcited cavity mode and ground-state atoms at equilibrium, to the Superradiant Phase (SP), with finite and constant cavity mode (and atomic) excitation. The critical coupling strength for the Dicke model transition, however, is of the order of the cavity mode and atomic transition frequencies, thus rendering it unfeasible in the optical domain with electric dipole transitions in real atoms.

More recently, though, a proposal by Dimer {\em et al}. \cite{dimer} showed that implementation of an effective Dicke model and demonstration of the phase transition was feasible in optical cavity QED through careful engineering of Raman transitions between hyperfine ground states of multilevel atoms. Moreover, their scheme allows for independent control over the rotating and counter-rotating terms in the (effective) atom-field interaction Hamiltonian -- a so-called ``unbalanced'' or ``generalized'' Dicke model -- thereby opening the door to possible new behavior. 

Experiments subsequently followed, utilizing this or similar kinds of Hamiltonian engineering to realize effective Dicke models in a variety of systems. These ranged from Bose-Einstein Condensates (BECs) or cold-atom clouds in optical cavities \cite{baum,klinder,zhiqiang1,zhiqiang2}, to spin-orbit-coupled BECs in harmonic traps \cite{hamner}, and trapped-ion arrays \cite{safavi}.

Uniquely, the experiment of \cite{zhiqiang1} in fact realized a spin-1 version of the Dicke model ({\em cf}. spin-1/2), using the $F=1$ hyperfine ground state of ${}^{87}$Rb and following a scheme similar to \cite{dimer}, as outlined in detail in \cite{masson1}. However, the experiment was performed in such a way that the phenomena observed only depended on the total collective spin length, rather than on the individual atomic spin. Nevertheless, 
their exploration of unbalanced coupling in a dissipative setting (due to cavity loss) revealed the existence of a third, Oscillatory Phase (OP), with the cavity photon number oscillating about a finite mean value in steady state. 
Spurred on by these results, Stitely {\em et al}. \cite{stitely} investigated the nonlinear, semiclassical (mean-field) ordinary differential equations (ODEs) describing the system in the thermodynamic limit ($N\to\infty$). They uncovered a rich variety of nonlinear phenomena and a complicated phase diagram for the system as function of the (unbalanced) atom-cavity coupling strengths. 

In comparison little work has been done on generalized Dicke models involving atoms with spin larger than 1/2. Masson {\em et al}. \cite{masson1,masson2} have investigated spin-1 versions of the Dicke model, but focus on specific applications in specific coupling regimes. Other investigations of three-level Dicke models \cite{skulte,fan,hayn1,hayn2,baksic,Chitra2022}, although more general in their choice of coupling, have considered closed systems, or specific energy-level structures, such as V or $\Lambda$ configurations. By constraining the level structure, these models potentially explore a smaller range of behavior and lose a degree of generality.

We therefore aim to generalize the spin-1 Dicke model by considering arbitrary coupling and level structures, in both dissipative and non-dissipative scenarios. We allow for unbalanced coupling by using the aforementioned scheme of \cite{zhiqiang1,masson1}, and for general atomic energy-level structures by introducing a quadratic Zeeman shift to the model. Control over this shift and the regular atomic splitting (which is already a part of the spin-1 Dicke model) effectively allows one to move the ground magnetic sublevels at will, and thereby engineer any level configuration. In the present investigation, we consider both a fully quantum-mechanical treatment with simplifying approximations, and an initial survey of the semiclassical behavior. For the former, we map the dynamics according to two behaviors (oscillatory or divergent), while in the latter we already find a plethora of dynamics and derive a rudimentary phase diagram. On the whole, our exploration provides an extension and adds generality to the spinor Dicke model, and consequently contributes to our understanding of this important and interesting physical system.

This work is structured as follows: in Section \ref{sec:model} we describe the physical system and model under consideration, show the Hamiltonian and master equation under various approximations, and introduce the quadratic Zeeman shift. Section \ref{sec:becs} demonstrates the connection between our model and that of single-mode spinor BECs under said approximations and in a particular parameter regime. In Section \ref{sec:closed} we investigate the closed system under the aforementioned approximations by characterizing the stability of operator equations of motion. In particular, we present the eigenvalues characterizing these equations, and use them to find regions in parameter space where the system diverges. In Section \ref{sec:open}, we extend this investigation to the open system by using a master equation method to obtain equations of motion for first and second order operator expectations. We analyze these equations as before, albeit numerically, and confirm that the addition of dissipation causes widespread divergence of the operator moments. To study the system's behaviour in these regions of divergence of the simpler model, we turn to a semiclassical description in Section \ref{sec:semiclassical}, first with no dissipation in Section \ref{sec:non_dissipative} and later with dissipation in Section \ref{sec:dissipative}.  We observe distinct behaviors in various, different parameter regimes, and create a rough map of these behaviours using a simple numerical scheme. Lastly, we conclude our findings and remark about future work in Section \ref{sec:conclusion}.

%: --------------------------------------------------------------
%:                  Model
% --------------------------------------------------------------
\section{Model}\label{sec:model}

\begin{figure}
    \centering
    \includegraphics{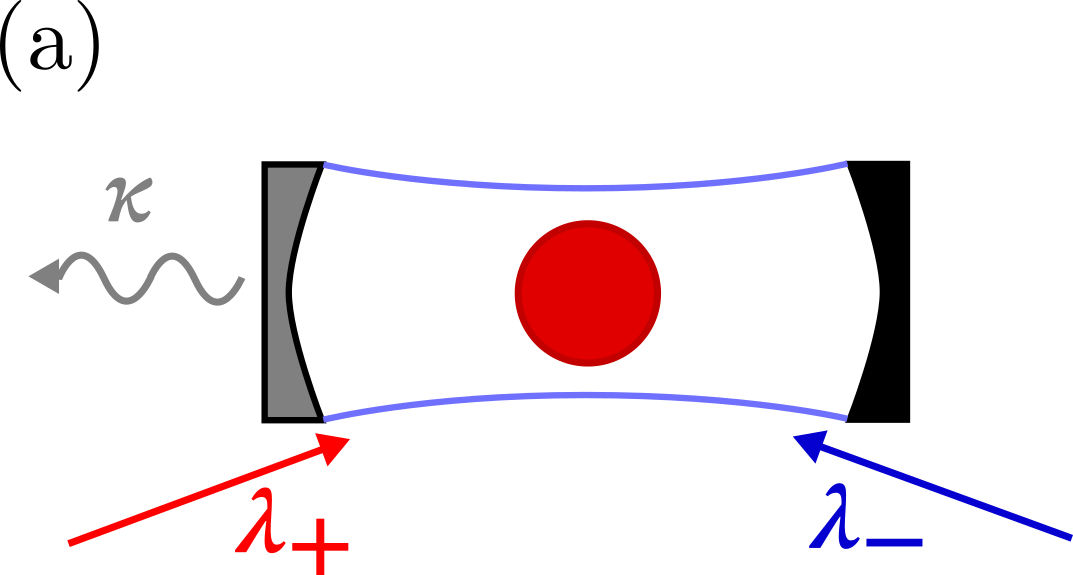}\vspace{1em}
    \includegraphics{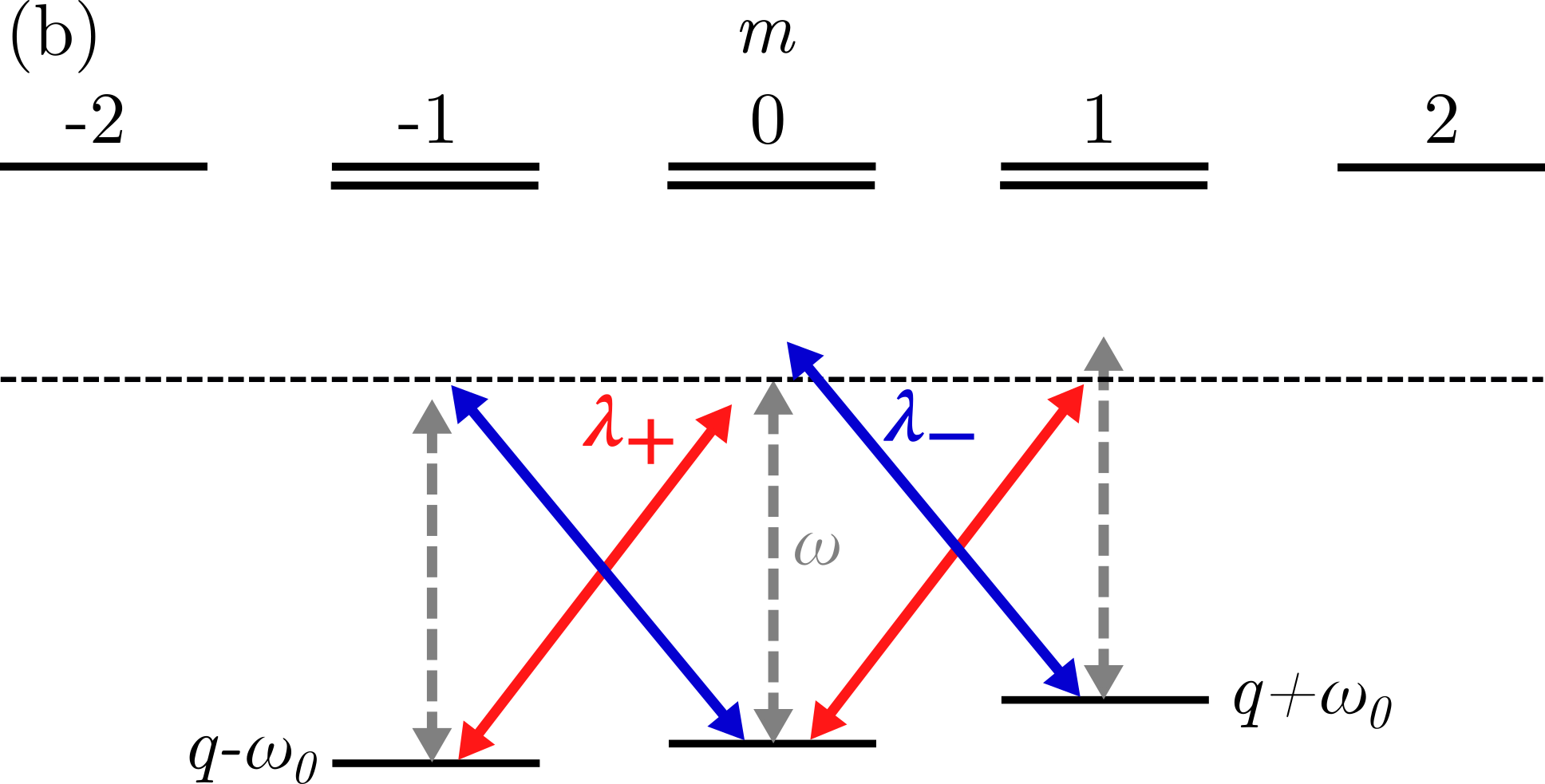}
    \caption{(a) Illustration of the physical system under consideration: an ensemble of spin-1 atoms in an optical cavity is driven by $\sigma_+$ (red arrow) and $\sigma_-$ (blue arrow) polarized light, and coupled to a single, $\pi$-polarized cavity mode (gray dashed line). 
    One mirror of the cavity is considered perfectly reflective, while the other is considered only partially reflective, with a cavity field decay rate $\kappa$. (b) Atomic level diagram of the implementation, where cavity-assisted Raman transitions occur between the three ground magnetic sublevels corresponding to $m=0,\pm1$. The $m=\pm1$ sublevels are shifted by $q\pm\omega_0$ from the $m=0$ sublevel, respectively.}
    \label{fig:setup}
\end{figure}

The physical system we wish to investigate is an ensemble of $N$ spin-1 atoms (e.g., ${}^{87}$Rb atoms in the $F=1$ ground hyperfine state) coupled collectively to a single mode of the electromagnetic field in an optical cavity, which we take to be linearly ($\pi$) polarized. The ensemble is additionally driven by two counter-propagating laser beams with $\sigma_-$ and $\sigma_+$ circular polarizations, respectively. As described by Masson {\em et al}. in Ref.~\cite{masson1}, and shown pictorially in Fig.~\ref{fig:setup}, atoms are able to absorb photons from the driving beams and emit into the cavity mode, or vice-versa, enabling transitions to magnetic sublevels of higher or lower magnetic number, depending on the absorbed photon. By tuning the cavity and laser frequencies far from the atomic dipole transition frequency, transitions into excited atomic states become off-resonant and the excited states are negligibly populated. Consequently, the cavity and laser fields predominantly drive Raman transitions between the ground state sublevels.

For very large detunings of the fields from the atomic transition frequency, such a system is described, in a frame rotating at the laser frequency, by the Dicke model, with Hamiltonian ($\hbar=1$)
\begin{eqnarray}\label{eq:dicke_ham}  
\Hh &=\omega\ah^\dag\ah+\om\Sh_z 
+\frac{\lambda_-}{\sqrt{2N}}(\ah\Sh_++\ah^\dag\Sh_-) \nonumber
\\
&+\frac{\lambda_+}{\sqrt{2N}}(\ah\Sh_-+\ah^\dag\Sh_+)\,,
\end{eqnarray}
where $\omega$ and $\omega_0$ are tunable, effective cavity and atomic frequencies, respectively, and $\lambda_\pm$ (given by Raman transition rates) are similarly tunable coupling constants for the co- and counter-rotating terms. Note that $\ah$ is the cavity mode annihilation operator, while $\Sh_z$ and $\Sh_\pm$ are collective atomic spin operators given by
\begin{equation}
    \Sh_{z,\pm}\equiv\sum_k^N\Sh_{z,\pm}^{(k)}\,,
\end{equation}
where $\Sh_{z,\pm}^{(k)}$ is the spin operator for the $k$-th atom.

We note that controlling $\omega_0$ allows for shifting of the $m=\pm1$ sublevels by equal and opposite amounts, i.e., $m=1$ is shifted up by $\omega_0$, while $m=-1$ is shifted down by $\omega_0$. Therefore, to allow for arbitrary sublevel structures we introduce a quadratic Zeeman shift to the model. Such a shift raises (or lowers) the energy of the $m=\pm1$ sublevels by an equal amount, which we specify through the parameter $q$, meaning the $m=\pm1$ sublevels are shifted, respectively, by $q\pm\omega_0$ in total.

More specifically, the quadratic shift is implemented via addition of the term
\begin{equation}\label{eq:Hqz}    
    \Hh_{QZ}=q\sum_{k=1}^N\Sh_z^{(k)2}
\end{equation}
to the Hamiltonian in Eq.~(\ref{eq:dicke_ham}), giving the total Hamiltonian
\begin{equation}\label{eq:total_ham}
    \Hh_T=\Hh+\Hh_{QZ}\,.
\end{equation}
{Note that $\Hh_{QZ}$ has no convenient representation in terms of the collective spin operators $\Sh_{z,\pm}$. Hence, unlike the standard Dicke model in Eq. (\ref{eq:dicke_ham}), the total Hamiltonian in Eq. (\ref{eq:total_ham}) cannot, for $q\neq0$, be written in terms of operators obeying SU(2) algebra. Instead, an SU(3) operator basis can be used, e.g., the Gell-Mann matrices \cite{Chitra2022}.

Physically realising such a quadratic Zeeman shift can be achieved with static magnetic fields, i.e., through the Zeeman effect, where $q$ would be positive and proportional to the square of the field strength. It could also be achieved with off-resonant laser fields to implement differential Stark shifts of the atomic levels, which potentially allows for negative $q$ \cite{you2017}.}

Lastly, we model dissipation due to cavity loss in the standard way via the master equation for the total system density operator $\roh$,
\begin{equation}\label{eq:full_me}
    \frac{d\roh}{dt}=-i[\Hh_T,\roh]+\kappa\mathcal{D}[\ah]\roh\,,
\end{equation}
where $\mathcal{D}$ is the superoperator defined by 
\begin{equation}
    \mathcal{D}[\Xh]\roh=2\Xh\roh\Xh^\dag-\Xh^\dag\Xh\roh-\roh\Xh^\dag\Xh\,.
\end{equation}

\subsection{Model with Adiabatic Elimination of the Cavity Mode}

In certain parts of this work we consider a regime where the cavity-assisted Raman transitions used to implement the model are themselves off-resonant, i.e., when $\omega\gg\{\lambda_\pm,\omega_0\}$. In this ``dispersive'' limit, the cavity mode is sparsely populated and effectively only mediates atom-atom interactions. We may then adiabatically eliminate the cavity mode by tracing over the cavity degrees of freedom and making standard approximations to arrive at the modified master equation,
\begin{equation}\label{eq:adiabatically_elim_me}
    \frac{d\roh_A}{dt}=-i[\Hh_S,\roh_A]+\frac{\kappa}{\kappa^2+\omega^2}\mathcal{D}[\Sh_\theta]\roh_A\,,
\end{equation}
where $\hat\rho_A$ is the atomic density operator, $\Hh_S$ is the adiabatically eliminated Hamiltonian,
\begin{align}
    \Hh_S&\equiv \Hh_{QZ}+\om\Sh_z-\frac{\omega}{\kappa^2+\omega^2}\Sh_\theta^\dag\Sh_\theta\nonumber\\
    &=\Hh_{QZ}+\om\Sh_z\nonumber\\
    &\hspace{1em}-\frac{\omega}{2N(\kappa^2+\omega^2)}\left[(\lambda_-+\lambda_+)^2\Sh_x^2\right.\nonumber\\
    &\hspace{8em}\left.+(\lambda_--\lambda_+)^2\Sh_y^2+(\lambda_-^2-\lambda_+^2)\Sh_z\right]\,,
\end{align}
and we have define
\begin{equation}
    \Sh_\theta\equiv\frac{1}{\sqrt{2N}}\left(\lambda_-\Sh_++\lambda_+\Sh_-\right)\,.
\end{equation}
The full derivation of this adiabatically eliminated Hamiltonian is detailed in the supplemental material of Ref.~\cite{masson1}.

We may also represent the atomic operators in terms of bosonic mode annihilation and creation operators (in the Jordan-Schwinger representation), denoted by $\bh_\pm$ for the $m=\pm1$ sublevels, and $\bh_0$ for the $m=0$ sublevel. These obey the regular annihilation and creation operator algebra, namely
\begin{align}
    [\bh_i,\bh_j]&=0\,, & [\bh_i,\bh_j^\dag]&=\delta_{ij}\,,
\end{align}
where $i,j=0,\pm$, and $\delta_{ij}$ is the Kronecker delta. {In this representation, summation over an arbitrary atomic operator $\Xh$ can be mapped according to \cite{schwinger}
\begin{equation}
    \sum_{k=1}^N\Xh^{(k)}\to\sum_{i,j}\bh^\dag_i\bh_j\bra{i}\Xh\ket{j}\,,
\end{equation}
where $i$ and $j$ have the same meaning as above, and $\ket{i}\equiv\ket{m=i}$. For example, $\Hh_{QZ}$ is mapped to
\begin{align}
    \Hh_{QZ}&\to q\sum_{i,j}\bh^\dag_i\bh_j\bra{i}\Sh_z^2\ket{j}\nonumber\\
    &=q\sum_{i,j}j^2\bh^\dag_i\bh_j\bra{i}\ket{j}\nonumber\\
    &=q(\bh^\dag_+\bh_++\bh^\dag_-\bh_-)\,.
\end{align}

One similarly finds that
\begin{align}
    \Sh_z&=\bh_+^\dag\bh_+-\bh_-^\dag\bh_-\,,\\
    \Sh_\pm&=\sqrt{2}(\bh_\pm^\dag\bh_0+\bh_0^\dag\bh_\mp)\label{eq:collective_spin_ops}\,.
\end{align}}

\subsection{Undepleted Mode Approximation}

When treating the system quantum-mechanically, we additionally specialize to the case where all atoms are initially prepared in the $m=0$ sublevel. If we further make the approximation that the ensemble contains a large number of atoms, i.e., essentially that $N\to\infty$, then we can assume that the $m=0$ sublevel remains macroscopically occupied throughout the evolution of the system. 
This allows us to replace the bosonic mode operator for the $m=0$ sublevel with a constant $c$-number, $\bh_0\rightarrow\sqrt{N}$, and consequently to write $\Sh_\pm$ as
\begin{equation}
    \Sh_\pm=\sqrt{2N}(\bh_\pm^\dag+\bh_\mp)\,.
\end{equation}

By further defining the operators
\begin{align}
    \Ah\equiv\bh_+^\dag+\bh_-\,, ~~~\Bh\equiv\bh_+^\dag-\bh_-\,,
\end{align}
which satisfy $[\Ah,\Ah^\dag ]=[\Bh,\Bh^\dag ]=[\Ah ,\Bh ]=0$, $[\Ah ,\Bh^\dag ]=[\Bh ,\Ah^\dag ]=-2$, 
we may rewrite the Hamiltonian as
\begin{align}\label{eq:full_ham}
    \Hh_T&=(q+\om)\frac{(\Ah+\Bh)(\Ah^\dag+\Bh^\dag)}{4}\nonumber\\
    &+(q-\om)\frac{(\Ah^\dag-\Bh^\dag)(\Ah-\Bh)}{4}\nonumber\\
    &-\Delta_+(\Ah^\dag+\Ah)^2-\Delta_-(\Ah^\dag-\Ah)^2\,,
\end{align}
where we have defined the constants
\begin{equation}
    \Delta_\pm=\frac{(\lambda_+\pm\lambda_-)^2}{4\omega}\,.
\end{equation}

Note that, implicit in the undepleted mode approximation is the assumption that the populations of the $m=\pm 1$ states satisfy $\langle\Nh_\pm\rangle\ll N$ at all times.
However, in certain regimes of parameter space, this condition is violated, which, as we shall see, manifests itself as divergence in the solutions of this simplified model.

%: --------------------------------------------------------------
%:                    Spinor BEC
% --------------------------------------------------------------
\section{Analogy with Spinor BEC Models}\label{sec:becs}

In the particular case where $\lambda_+=0$, the adiabatically-eliminated model described above has already been investigated to some extent by Masson {\em et al}. in Refs.~\cite{masson1,masson2}. In this case, the cavity-mediated atomic interactions can emulate collisional interactions in a single-mode spinor BEC, which, with the inclusion of the quadratic Zeeman shift, are described by a Hamiltonian of the general form
\begin{equation}
    \Hh=\frac{\Lambda}{N}\Sh^2+q(\Nh_++\Nh_-)\,,
\end{equation}
where {$\Nh_\pm\equiv\bh^\dag_\pm\bh_\pm$, and} the signs and relative strengths of the parameters $\Lambda$ and $q$ determine the ground state phases of the system. 

Effective realization of this Hamiltonian was shown to be possible based upon the spinor Dicke model presented here, together with the possibility of spin-nematic squeezing \cite{masson1}, which has indeed been demonstrated experimentally \cite{Davis2019}. Preparation of the atomic ensemble in the spin singlet state, where the collective spin is zero and atom-atom entanglement is strong, has also been proposed \cite{masson2}, highlighting the potential utility of the model.

{An extensive body of research also exists on the spinor BEC side. There have been several investigations of spin-nematic squeezing in this context \cite{duan2002,hamley2012,hoang2013}, as well as spin mixing dynamics \cite{law1998,pu1999}. More specifically, Pu \textit{et al.} \cite{pu1999} used an operator mapping from field operators to annihilation and creation operators, and found a range of dynamical behaviours including persistent oscillations.

In a similar vein, recent work by Evrard \textit{et al.} \cite{evrard1,evrard2} explored the system dynamics both semiclassically (i.e., via a mean-field description) and quantum mechanically.} They observed two qualitatively different dynamics, based on the value of $q$: for large $q$ the populations of the $m=\pm1$ sublevels oscillated with small amplitudes, while for small $q$ there was significant depletion of the $m=0$ sublevel and the dynamics eventually relaxed towards a constant steady-state. These results were in good agreement with numerical solutions of the Schr\"odinger equation, but in various degrees of agreement with the semiclassical description, depending on initial seeding of the $m=\pm1$ sublevels. Larger seeds were less affected by quantum fluctuations, and thus better approximated by the semiclassical description, which neglects these fluctuations entirely.

To test the analogy between our model and the single-mode BEC system, we follow Masson {\em et al}. and set $\lambda_+=0$ (or $\lambda_-=0$) and $\lambda_-=\lambda$ (or $\lambda_+=\lambda$), with the cavity mode adiabatically eliminated \cite{masson1,masson2}. 
For simplicity and to give the best comparison,
we set $\kappa=0$ in our model and consider Hamiltonian dynamics only. This gives the Hamiltonian
\begin{align}
    \Hh_S&=q(\Nh_++\Nh_-)+\om\Sh_z-\frac{\lambda^2}{2N\omega}\left( \Sh_x^2+\Sh_y^2\right)\nonumber\\\nonumber\\
    &=(q+\omega_0)\bh_+^\dag\bh_++(q-\omega_0)\bh_-^\dag\bh_-\nonumber\\&\hspace{1em}-\Lambda(\bh_++\bh_-^\dag)(\bh_+^\dag+\bh_-)\label{eq:TMS}
    \\\nonumber\\
    &=(q+\om)\frac{(\Ah+\Bh)(\Ah^\dag+\Bh^\dag)}{4}\nonumber\\&\hspace{1em}+(q-\om)\frac{(\Ah^\dag-\Bh^\dag)(\Ah-\Bh)}{4}-\Lambda\Ah^\dag\Ah\,,
\end{align}
where we defined $\Lambda\equiv\lambda^2/\omega$, and the term $-\Lambda\Sh_z$ has been incorporated into the term $\omega_0\Sh_z$.

{The form ({\ref{eq:TMS}}) highlights the connection to two-mode squeezing, as shown by Masson \textit{et al.} \cite{masson1}. Here, however, we focus instead on the general dynamics and their connection with the aforementioned spinor BEC work. To this end, we find the Heisenberg equations of motion for $\Ah$ and $\Bh$:}
\begin{equation}
   \frac{d}{dt}\begin{bmatrix}\Ah\\\Bh\end{bmatrix}=i\begin{bmatrix}\om & q\\q-2\Lambda & \om\end{bmatrix}\begin{bmatrix}\Ah\\\Bh\end{bmatrix}\, .
\end{equation}

The qualitative behavior of solutions to this system, and by extension the qualitative behavior of the atomic dynamics, are determined by the eigenvalues of the matrix appearing in the above equation. If any of said eigenvalues have a positive real part, then solutions will diverge, rendering the model nonphysical, at least on long timescales, since atomic populations are clearly not permitted to grow indefinitely. Otherwise, solutions will be superpositions of complex exponentials and will exhibit simple oscillations.

\begin{figure}
    \centering
    \includegraphics[width=0.49\textwidth]{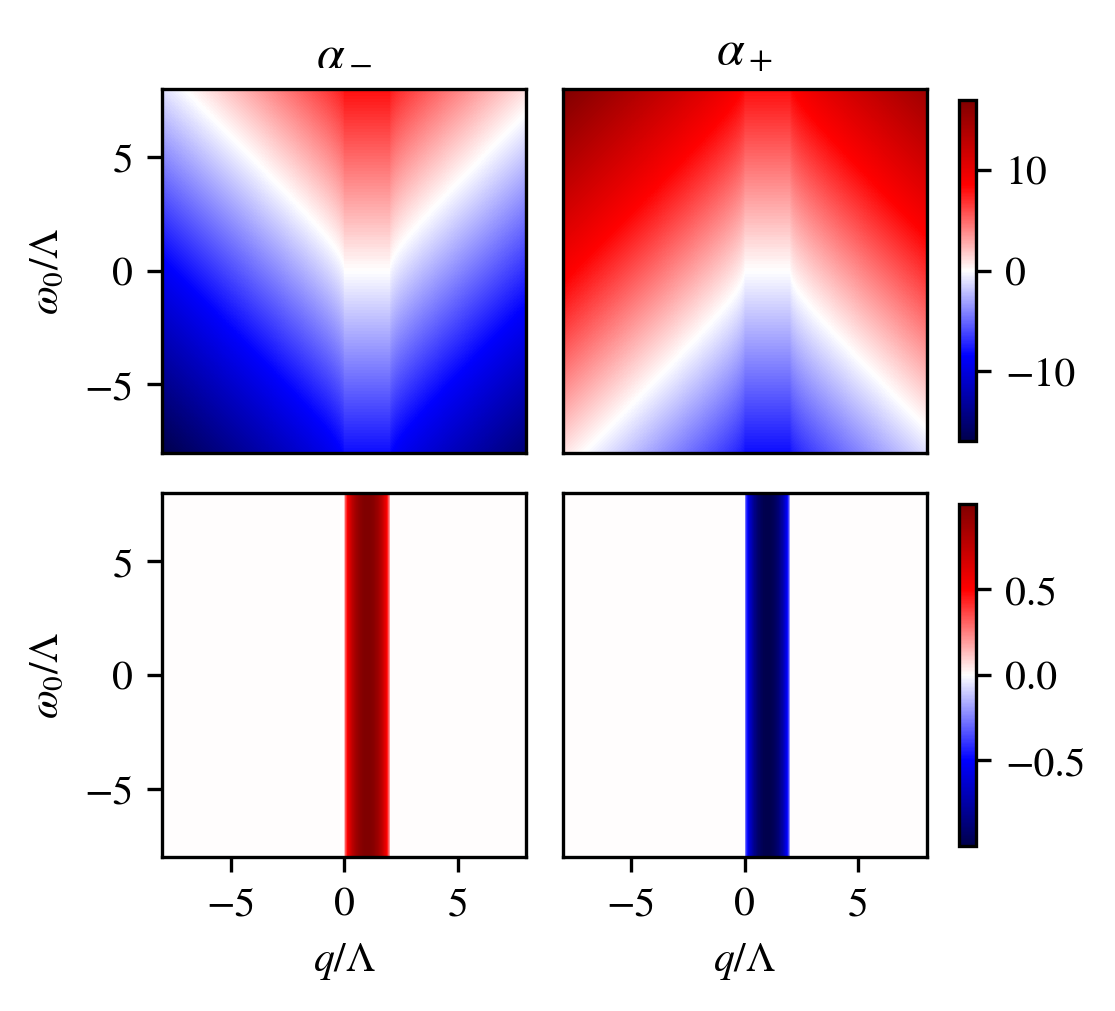}
    \caption{Imaginary (top row) and real (bottom row) parts of $\alpha_-$ (left two plots) and $\alpha_+$ (right two plots) as functions of $q$ and $\om$. The colourbars apply to each row.}
    \label{fig:lp=0_landscape}
\end{figure}

The eigenvalues are
\begin{equation}
    \alpha_\pm=i\om\pm i\sqrt{q(q-2\Lambda)}\,.
\end{equation}
We will assume that $\Lambda>0$.
If $0< q<2\Lambda$, one eigenvalue will have positive real part, otherwise both eigenvalues have zero real part, as shown in Fig.~\ref{fig:lp=0_landscape}, where we plot the real and imaginary parts of $\alpha_\pm$ as functions of $q$ and $\omega_0$. Their behavior is in agreement with the aforementioned experimental results of Evrard {\em et al}. \cite{evrard1}, where our analogous oscillatory behavior occurs when $q>2\Lambda$ (or $q<0$). However, our model diverges and becomes nonphysical when $0<q<2\Lambda$, and does not replicate their relaxation dynamics results; these indeed correspond to major depletion of the $m=0$ sublevel, which we have neglected by making the undepleted mode approximation. 

Explicitly, the solutions for the populations of the $m=\pm 1$ sublevels that we obtain from our model, with initial conditions $\langle \Nh_\pm (t=0)\rangle =0$, are
\begin{equation}
\langle \Nh_\pm (t)\rangle = \frac{\Lambda^2}{q(q-2\Lambda)} \, \sin^2 \left( \sqrt{q(q-2\Lambda)}t\right)
\end{equation}
for $q>2\Lambda$ (or $q<0$), and
\begin{equation}
\langle \Nh_\pm (t)\rangle = \frac{\Lambda^2}{q(2\Lambda -q)} \, \sinh^2 \left( \sqrt{q(2\Lambda -q)}t\right)
\end{equation}
for $0<q<2\Lambda$.\\

%: --------------------------------------------------------------
%:                  Closed Quantum
% --------------------------------------------------------------
\section{Generalized Dicke Model: Quantum Mechanical Treatment}
\subsection{Closed System}\label{sec:closed}

Unlike the single-mode spinor BEC systems, our engineered Dicke model has additional freedom with regards to tuning parameters and interactions. For example, if we set $\lambda_+=\lambda_-=\lambda$ then atom-atom interactions are governed by a term solely proportional to $\Sh_x^2$ (rather than $\Sh^2$), which, as we shall see, leads to significant changes in behavior. 
In this subsection we still consider a closed system, where operators evolve subject to the Hamiltonian (\ref{eq:full_ham}), to maintain the single-mode spinor BEC analogy. We find the equations of motion for $\Ah$, $\Bh$, and their adjoints in the fully general case are given by
\begin{widetext}
\begin{equation}\label{eq:closed_eoms}
    \frac{d}{dt}\begin{bmatrix}\Ah\\\Ah^\dag\\\Bh\\\Bh^\dag\end{bmatrix}=i\begin{bmatrix}
    \om & 0 & q & 0\\
    0 & -\om & 0 & -q\\
    q-4(\Delta_++\Delta_-) & -4(\Delta_+-\Delta_-) & \om & 0\\
    4(\Delta_+-\Delta_-) & -q+4(\Delta_++\Delta_-) & 0 & -\om
    \end{bmatrix}\begin{bmatrix}\Ah\\\Ah^\dag\\\Bh\\\Bh^\dag\end{bmatrix}\,.
\end{equation}
The eigenvalues of the above matrix are found to be
\begin{equation}\label{eq:closed_evals}
    \alpha=\pm i\sqrt{ \pm2\sqrt{q\left\{4(\Delta_+-\Delta_-)^2q+\om^2\left[q-4(\Delta_++\Delta_-)\right]\right\}}+\om^2+q\left[q-4(\Delta_++\Delta_-)\right]}\,,
\end{equation}
\end{widetext}
where each choice of $\pm$ denotes a different eigenvalue. Two eigenvalues are negatives of the other two, and, consequently, if any eigenvalue has a nonzero real part there must be an eigenvalue with a positive real part, again leading to divergence. 
Hence, we need only consider two linearly independent eigenvalues to characterize the behavior of the system. We choose these to have the positive outer sign and opposite inner signs in (\ref{eq:closed_evals}), and we denote these by $\alpha_\pm$ according to said inner sign.

\begin{figure}
    \centering
    \includegraphics[width=0.49\textwidth]{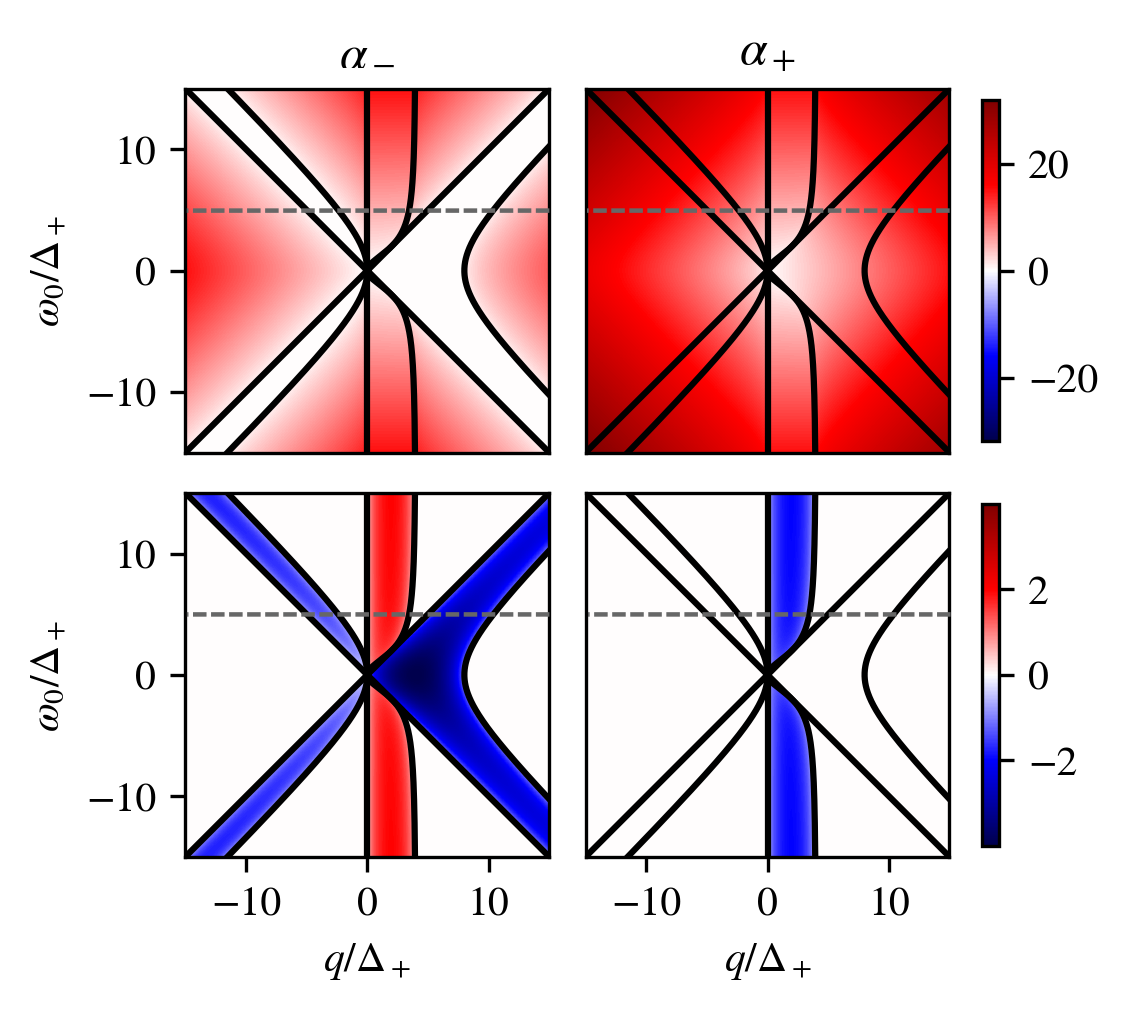}
    \caption{Imaginary (top row) and real (bottom row) parts of $\alpha_-$ (left two plots) and $\alpha_+$ (right two plots) as functions of $q$ and $\om$ for $\Delta_-=0$. Boundaries separating the regions of divergence from regions of oscillation, as given by Eqs.~(\ref{eq:closed_boundaries_1},\ref{eq:closed_boundaries_2}), are plotted in black. {The dashed gray lines represent a constant slice through the eigenvalue landscape for $\omega_0=5\Delta_+$.} The colour bars apply to each row.}
    \label{fig:Dm=0_landscape}
\end{figure}

Fig.~\ref{fig:Dm=0_landscape} shows the real and imaginary parts of $\alpha_\pm$ as functions of $q$ and $\omega_0$, for balanced coupling (i.e., $\lambda_+=\lambda_-$). We see distinct regions where $\alpha_-$ has a nonzero real part, which also include the regions where $\alpha_+$ has a nonzero real part. The system's divergence is then conditional purely on the real part of $\alpha_-$, a condition which holds for all $\Delta_-$.

We can additionally find the boundaries separating the regions of divergence and oscillation, which occur when $\Re e(\alpha_-)=0$. This condition is satisfied in the specific case where $\alpha_-=0$, when the determinant of the matrix in Eq.~(\ref{eq:closed_eoms}) is necessarily zero. Although the converse is not generally true, i.e., a zero determinant does not imply $\alpha_-=0$, we may nevertheless find the determinant's roots and choose those that correspond to the desired boundaries.
The determinant is
\begin{equation}\label{eq:closed_det}
    D=\left[\om^2-q(q-8\Delta_-)\right]\left[\om^2-q(q-8\Delta_+)\right]\,,
\end{equation}
and its four roots are
\begin{equation}\label{eq:closed_boundaries_1}
    q=4\Delta_+\pm\sqrt{16\Delta_+^2+\om^2}\,,~~4\Delta_-\pm\sqrt{16\Delta_-^2+\om^2}\,.
\end{equation}

The remaining boundaries are given by the roots of the nested square root in (\ref{eq:closed_evals}), when $\alpha_\pm$ both become purely imaginary, and are found to be
\begin{equation}\label{eq:closed_boundaries_2}
    q=0\,,~\frac{4(\Delta_++\Delta_-)\om^2}{\om^2+4(\Delta_+-\Delta_-)^2}\,.
\end{equation}
These are all plotted over the eigenvalue landscapes in Fig.~\ref{fig:Dm=0_landscape}, where we see they indeed define all of the boundaries.

\begin{figure}
    \centering
\includegraphics{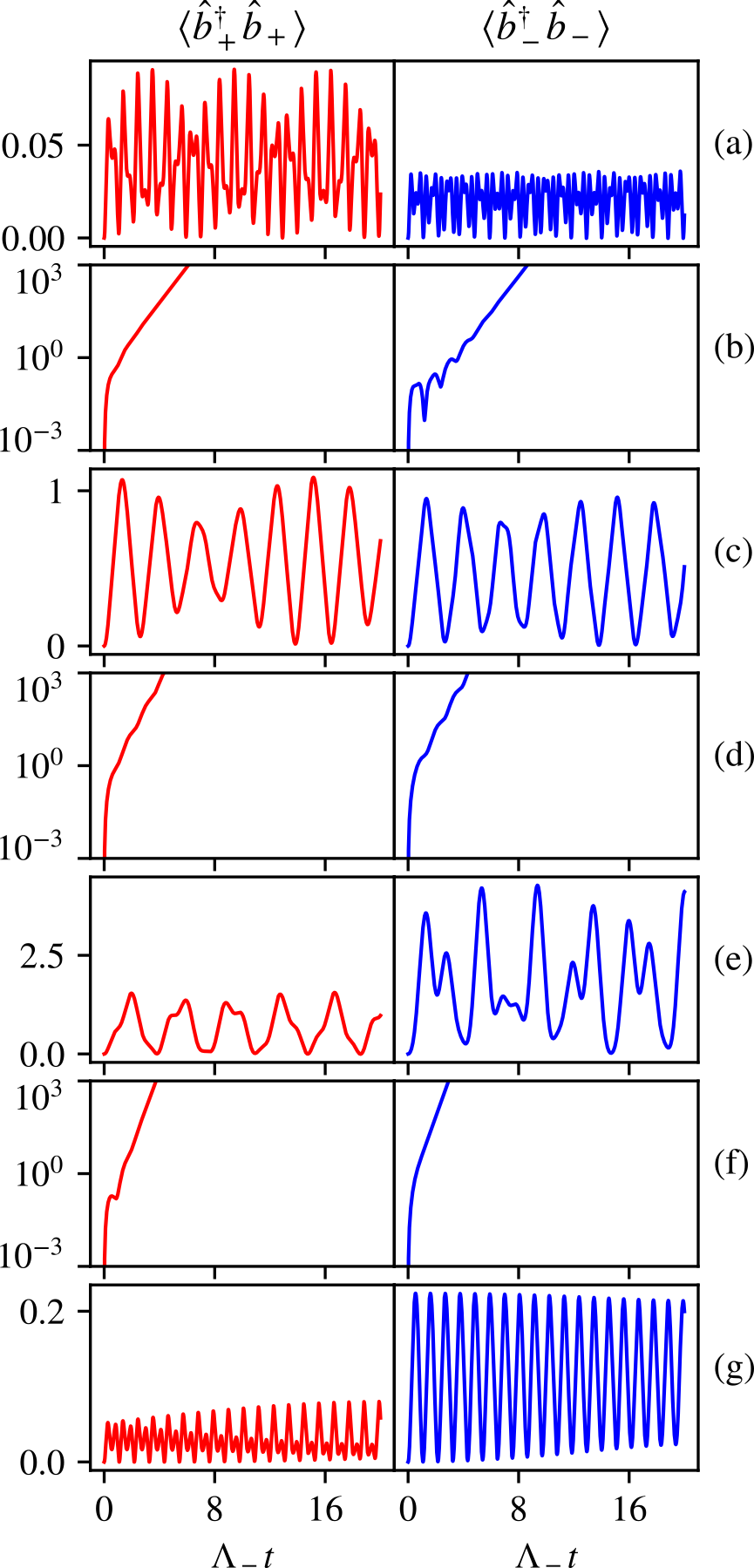}
    \caption{Populations of the $m=\pm 1$ sublevels as functions of time for $\Delta_-=0$, $\omega_0=5\Delta_+$, and $q/\Delta_+=-10$ (a), $-4$ (b), $-1$ (c), $2$ (d), $4$ (e), $8$ (f), $14$ (g). Note that rows (b), (d), and (f) are shown on a log scale.
    }
    \label{fig:populations_pm1}
\end{figure}

The eigenvalue landscape, or ``phase diagram'' for this generalized model clearly has much more structure than that of the effective BEC model of the previous section. There is now a strong dependence on both $\omega_0$ and $q$, with multiple regions of stability and instability as a function of either parameter. The behavior of the populations of the $m=\pm 1$ sublevels also display quite distinct behaviors in the different regions, as demonstrated in Fig.~\ref{fig:populations_pm1}, where the populations are plotted as a function of time for each of the seven distinct regions of stability or instability. {These are associated with the ``slice'' of the eigenvalue landscape from Fig. \ref{fig:Dm=0_landscape} through the line of constant $\omega_0=5\Delta_+$, which crosses through each of the seven aforementioned regions. In each case from Fig.~\ref{fig:populations_pm1}, the population time series were obtained by numerically integrating the Heisenberg equations of motion for the relevant second-order operator expectations, using a standard 4th-order Runge-Kutta method with a sufficiently small stepsize. These equations of motion are given in Appendix \ref{appdx:mom_eqns} and are discussed further in the case of the general, open system in Sec. \ref{sec:open}. In principle, however, this is not required; given the system in Eq.~(\ref{eq:closed_eoms}) is linear, one could solve it exactly. The solutions would be superpositions of complex exponentials, whose oscillation frequencies are given by the imaginary parts of the eigenvalues in Eq.~(\ref{eq:closed_evals}). When none of those eigenvalues have a positive real part, as in Figs.~\ref{fig:populations_pm1}(a), (c), (e), and (g), the solutions oscillate ad infinitum. Otherwise, as in Figs.~\ref{fig:populations_pm1}(b), (d), and (f), the solutions diverge. Hence, although the behaviours in Fig. \ref{fig:populations_pm1} may appear complicated, they are simple combinations of finitely many oscillation frequencies or exponential increases.}

\begin{figure}[t]
    \centering   \includegraphics{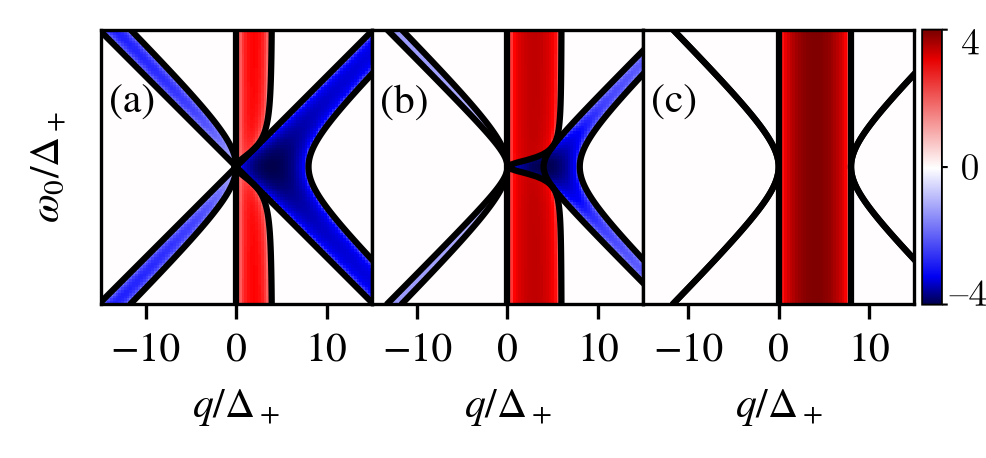}
    \caption{Real parts of $\alpha_-$ and divergence boundaries as functions of $q$ and $\om$ for $\Delta_-/\Delta_+= $ 0 (a), $0.5$ (b), and $1$ (c).}
    \label{fig:landscapes}
\end{figure}

Figure ~\ref{fig:landscapes} shows the real parts of $\alpha_-$ and the divergence boundaries for various choices of $\Delta_-$. We see a gradual transition from the landscape in Fig.~\ref{fig:Dm=0_landscape} to a single, vertical band as $\Delta_-$ tends to $\Delta_+$ (i.e., as $\lambda_+\rightarrow 0$), as seen in the previous section. {Most importantly, as evident in panel (b), the qualitative structure of the eigenvalue landscape survives through this transition. }

%: --------------------------------------------------------------
%:                  Open Quantum
% --------------------------------------------------------------
\subsection{Open System}\label{sec:open}

Given our system is considered in a quantum optical setting, in order to realistically model it we must take dissipation into account. Doing so will give us a more complete description of the dynamics and thereby a possible explanation for the behavior of the system in the regions of divergence from the previous section.

We begin by considering the master equation in Eq.~(\ref{eq:adiabatically_elim_me}) to find the equations of motion for various operator moments. For an arbitrary operator $\Xh$, we have
\begin{align}\label{eq:op_exp}
    \frac{d\expv{\Xh}}{dt}&=\Tr\left\{\Xh\frac{d\roh_A}{dt}\right\}\nonumber\\
    &=-i\,\Tr\left\{\Xh[\Hh_S,\roh_A]\right\}+\frac{\kappa}{\kappa^2+\omega^2}\Tr\left\{\Xh\mathcal{D}[\Sh_\theta]\roh_A\right\}\nonumber\\
    &=-i\expv{[\Xh,\Hh_S]}
    \nonumber\\
    &~~~~~+\frac{\kappa}{\kappa^2+\omega^2}\left(\expv{\Sh_\theta^\dag[\Xh,\Sh_\theta]}-\expv{[\Xh,\Sh_\theta^\dag]\Sh_\theta}\right) .
\end{align}
Given $\Hh_S$ can be written in terms of just $\Sh_\theta$ and $\Nh_\pm$, in order to find the equation of motion for $\expv{\Xh}$ one simply needs to find the commutators $[\Xh,\Sh_\theta]$, $[\Xh,\Sh_\theta^\dag]$, and $[\Xh,\Nh_\pm]$. Doing so for $\expv{\Ah}$, $\expv{\Bh}$, and their conjugates, one finds
\begin{widetext}
\begin{equation}\label{eq:open_eoms}
    \frac{d}{dt}\begin{bmatrix}\expv{\Ah} \\ \expv{\Ah^\dag} \\ \expv{\Bh} \\ \expv{\Bh^\dag}\end{bmatrix}=\begin{bmatrix}
    i\om & 0 & iq & 0\\
    0 & -i\om & 0 & -iq\\
    iq+2[\Gamma_+-\Gamma_--i(\Lambda_++\Lambda_-)] & -4i\sqrt{\Lambda_+\Lambda_-} & i\om & 0\\
    4i\sqrt{\Lambda_+\Lambda_-} & -iq+2[\Gamma_+-\Gamma_-+i(\Lambda_++\Lambda_-)] & 0 & -i\om
    \end{bmatrix}\begin{bmatrix}\expv{\Ah} \\ \expv{\Ah^\dag} \\ \expv{\Bh} \\ \expv{\Bh^\dag}\end{bmatrix}\,,
\end{equation}
\end{widetext}
where we define the constants
\begin{align}
    \Lambda_\pm&\equiv\frac{\omega\lambda_\pm^2}{\kappa^2+\omega^2} & \text{and} && \Gamma_\pm&\equiv\frac{\kappa\lambda_\pm^2}{\kappa^2+\omega^2}\,.
\end{align}

We may now perform a similar eigenvalue analysis as in the previous section to find the regions in parameter space where this system diverges (or not). One should note that this is a system of moment equations, not operator equations, meaning the stability of (\ref{eq:open_eoms}) does not directly determine the stability of higher-order moments. Nevertheless, since the higher-order moments inherently involve the same operators as the first-order moments, we expect them to diverge or oscillate in the same regions of parameter space.

We should also note that although the matrix in (\ref{eq:open_eoms}) is not significantly different from that in (\ref{eq:closed_eoms}), its eigenvalues have significantly more complicated forms. We therefore evaluate them numerically at each point in parameter space and extract the maximal real part from them, which is sufficient to determine the stability.

\begin{figure}
    \centering
\includegraphics{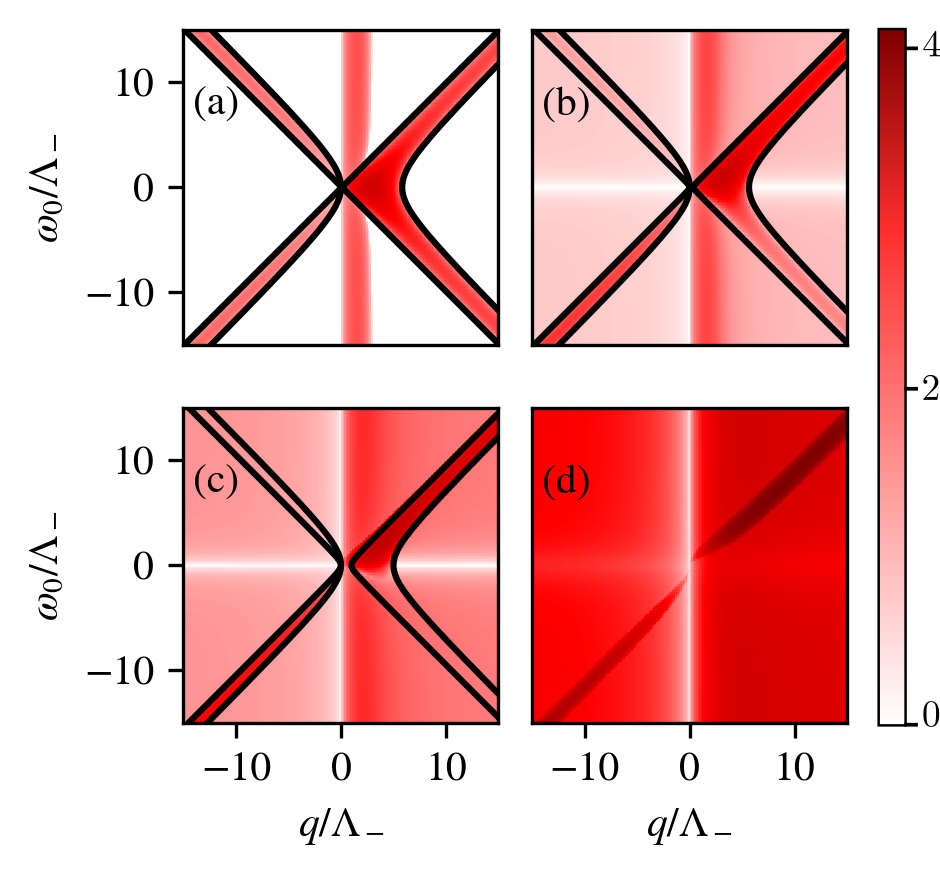}
    \caption{Real-part eigenvalue landscapes for $\Lambda_+=0.5\Lambda_-$ and increasing $\kappa$, as well as the roots of the determinant as given in Eq. (\ref{eq:open_roots}). The values of $\kappa/\Lambda_-$ used are 0 (a), 1 (b), 2 (c), and 5 (d)}
    \label{fig:open_landscapes}
\end{figure}

The determinant of the matrix is however comparatively simple, and given by
\begin{align}\label{eq:open_det}
    D&=\left\{\om^2-q\left[q-2(\Lambda_++\Lambda_-)(1+K)\right]\right\}\nonumber\\
    &\hspace{2em}\times\left\{\om^2-q\left[q-2(\Lambda_++\Lambda_-)(1-K)\right]\right\}\,,
\end{align}
where we define
\begin{equation}
    K\equiv\sqrt{1-\left(\frac{\kappa^2}{\omega^2}+1\right)\left(\frac{\Lambda_+-\Lambda_-}{\Lambda_++\Lambda_-}\right)^2}\,.
\end{equation}
The roots of this determinant are
\begin{align}\label{eq:open_roots}
    q&=(\Lambda_++\Lambda_-)(1+K)
    \nonumber\\
&~~~~\pm\sqrt{(\Lambda_++\Lambda_-)^2(1+K)^2+\om^2}\,,
\end{align}
and
\begin{align}\label{eq:open_roots}
    q&=(\Lambda_++\Lambda_-)(1-K)
    \nonumber\\
    &~~~~\pm\sqrt{(\Lambda_++\Lambda_-)^2(1-K)^2+\om^2}\,,
\end{align}
 which, given the analysis in the previous section, we expect will correspond to boundaries of regions of divergence. In Fig. \ref{fig:open_landscapes} we plot these roots alongside our numerical eigenvalue landscapes for various values of $\kappa$, where we see this is actually not the case.

The addition of a nonzero $\kappa$ adds a finite background to the landscape, which increases in magnitude with $\kappa$ and eventually dominates over the landscape. The boundaries between oscillation and divergence consequently become blurred, and the roots in (\ref{eq:open_roots}) just appear to separate regions of fast divergence (i.e., large real part) and slow divergence (i.e., small real part). In fact, the roots do not separate these regions perfectly, especially for small $q$ and $\omega_0$, but instead only give their approximate outline.

Lastly, we see that the boundaries disappear when $\kappa$ becomes sufficiently large. This occurs when $K$, and consequently the roots, acquire a nonzero imaginary part, or, more specifically, when $\kappa$ reaches the value
\begin{equation}
    \frac{\kappa}{\omega}=\frac{2\sqrt{\Lambda_+\Lambda_-}}{|\Lambda_+-\Lambda_-|}\,.
\end{equation}

\noindent Beyond this value of $\kappa$ the landscapes approach a mostly uniform shape, tending to a constant value with large $q$, and having no $\omega_0$ dependence.

\subsubsection{Population Dynamics}\label{sec:pop_dyn}

One should note that balanced coupling (i.e., $\lambda_+=\lambda_-$) was not mentioned in the eigenvalue analysis above. Indeed, only differences between $\Gamma_+$ and $\Gamma_-$ appear in Eq.~(\ref{eq:open_eoms}), so any dissipative terms are canceled when the coupling is balanced. This is not to say dissipation does not affect the dynamics; one simply needs to consider second-order moments. These are of additional interest, since they allow us to evaluate the atomic populations and thereby get a better understanding of the system's behavior for varying parameter choices.

To find a closed set of equations of motion for the second-order moments, we may use Eq.~ (\ref{eq:op_exp}) in combination with the parameter and operator exchange symmetry of our model. That is to say, under the transformation
\begin{equation}
    \mathcal{T}:(\omega_0,\lambda_+,\lambda_-,\bh_+,\bh_-)\to(-\omega_0,\lambda_-,\lambda_+,\bh_-,\bh_+)\,,
\end{equation}
both the master equation and Hamiltonian in (\ref{eq:total_ham}) and (\ref{eq:adiabatically_elim_me}) remain invariant. We can therefore obtain the equation of motion for $\expv{\bh_-}$ by applying $\mathcal{T}$ to the equation of motion for $\expv{\bh_+}$, and similarly with higher order moments such as $\expv{\bh_-^2}$ or $\expv{\bh_+\bh_-^\dag}$. If we further use conjugation to obtain the equations of motion for adjoint operator moments (e.g., $\expv{\bh_+^\dag}$ from $\expv{\bh_+}$), we only need to find four second order moment equations to find the remaining six.

\begin{figure}
    \centering
\includegraphics{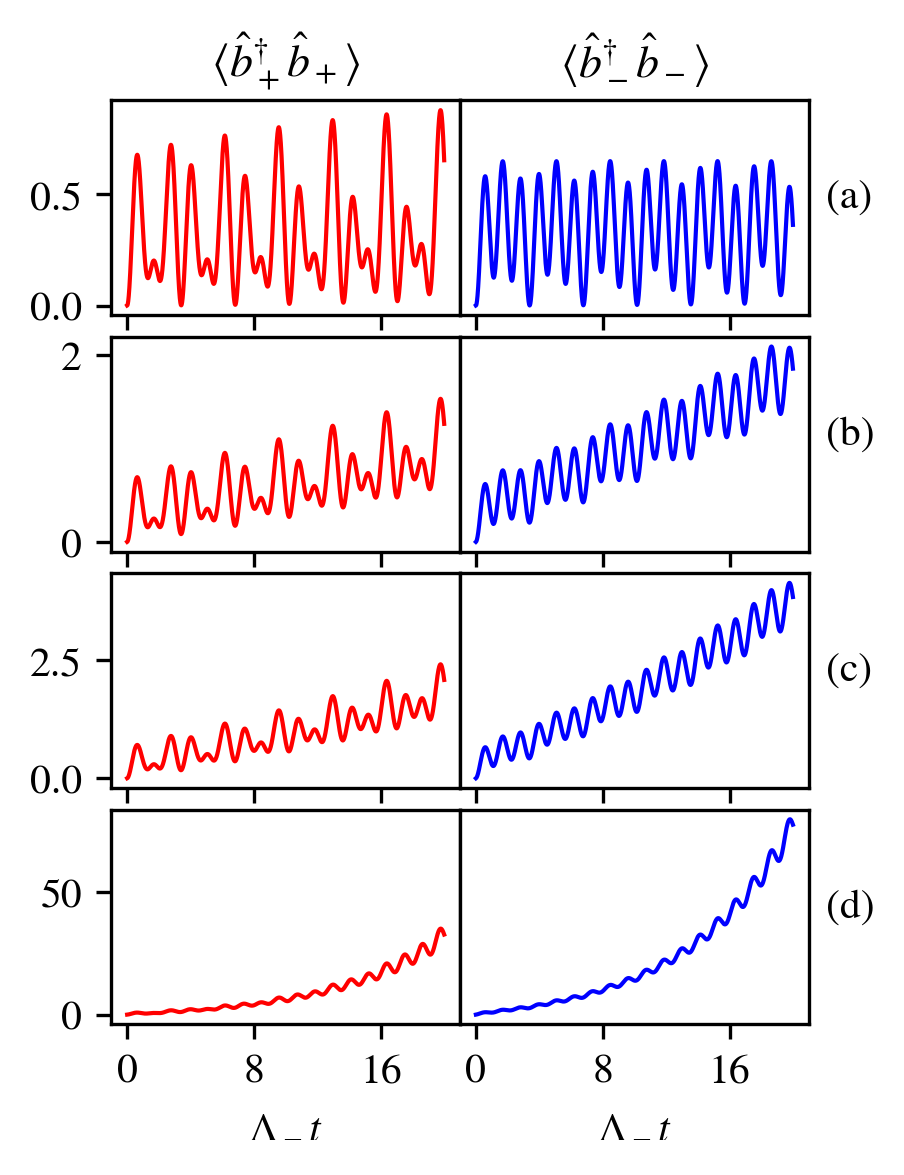}
    \caption{Population expectations for $\Lambda_+=\omega_0=0.5\Lambda_-$, $q=-\Lambda_-$, and increasing values of $\kappa$: row (a) corresponds to $\kappa=0$, (b) to $\kappa=0.05$, (c) to $\kappa=0.1$, and (d) to $\kappa=0.5$. The initial conditions for all of these correspond to initially unpopulated $m=\pm1$ sublevels.}
    \label{fig:open_pop_exps}
\end{figure}

These are all given in Appendix \ref{appdx:mom_eqns} and can be numerically integrated to find the population dynamics. One should additionally note that they all contain inhomogeneous terms, corresponding to the averaged effects of quantum fluctuations, meaning an initial state with no atoms in the $m=\pm1$ sublevels is nonstationary. Fig.~\ref{fig:open_pop_exps} shows the results of four numerical integrations for increasing values of $\kappa$ and a particular choice of $\Lambda_+$, $\omega_0$, and $q$, which for $\kappa=0$ corresponded to a region of oscillation. We see that for $\kappa>0$ the previous oscillation is superimposed over exponential growth, which becomes more rapid with increasing $\kappa$, in agreement with the eigenvalue analysis.

For sufficiently small $\kappa$ and over sufficiently small timescales, the sublevels still remain sparsely populated, meaning our undepleted $m=0$ mode approximation should still hold to a certain extent and these results should provide a reasonably accurate picture of the dynamics. However, for all realistic scenarios where $\kappa>0$ our approximation will eventually break down as depletion of the $m=0$ mode becomes significant. In fact, as we will see, finite $\kappa$, in combination with adiabatic elimination of the cavity mode, necessarily leads to an unphysical irreversibility and subsequent divergence in the model. 

Therefore, in order to investigate the system more deeply, we must take not only the dynamics of the $m=0$ mode into account, but also that of the cavity. To do so, we turn to a semiclassical analysis in the next section.

%: --------------------------------------------------------------
%:                  Semiclassical
% --------------------------------------------------------------
\section{Semiclassical Analysis}\label{sec:semiclassical}
\subsection{Non-Dissipative Dynamics}\label{sec:non_dissipative}

By relaxing the undepleted mode approximation, we inherently need to work with the full form of the collective spin operators, given by Eq. (\ref{eq:collective_spin_ops}). Given products of these operators appear in the Hamiltonian, inclusion of the $m=0$ mode leads to a non-quadratic Hamiltonian, which leads to a coupling between increasingly higher-order moments in the equations of motion, which never form a closed set of equations. For example, first order moment equations couple to third order moments, which themselves couple to fifth order moments, etc.

We therefore take the thermodynamic limit, where $N\to\infty$ and the non-commuting operators can be replaced with commuting $c$-numbers. This allows us to factorize high order moments into products of first order moments, and thus arrive at a closed set of semiclassical nonlinear equations for the first order moments. We define the $c$-numbers
\begin{align}
    \beta_\pm&=\frac{\expv{\bh_\pm}}{\sqrt{2N}}\,, & \beta_0&=\frac{\expv{\bh_0}}{\sqrt{2N}}\,,
\end{align}
and use the Hamiltonian and master equation in Eqs.~ (\ref{eq:total_ham}) and (\ref{eq:adiabatically_elim_me}) to arrive at the (closed) set of equations
\begin{align}
    \frac{d\beta_+}{dt}&=(i\Lambda_--i(q+\om)-\Gamma_-)\beta_+\nonumber\\
    &\hspace{1em}+(i\sqrt{\Lambda_+\Lambda_-}-\sqrt{\Gamma_+\Gamma_-})\beta_-\nonumber\\
    &\hspace{1em}+(i(\Lambda_++\Lambda_-)+\Gamma_+-\Gamma_-)(\beta_+|\beta_0|^2+\beta_-^*\beta_0^2)\nonumber\\
    &\hspace{1em}+2i\sqrt{\Lambda_+\Lambda_-}(\beta_-|\beta_0|^2+\beta_+^*\beta_0^2)\,,\label{eq:beta_+_eom}\\\nonumber\\
    \frac{d\beta_0}{dt}&=i(\Lambda_++\Lambda_-)(2\beta_+\beta_-\beta_0^*+(|\beta_+|^2+|\beta_-|^2)\beta_0+\beta_0)\nonumber\\
    &\hspace{1em} +2i\sqrt{\Lambda_+\Lambda_-}(\beta_+^2\beta_0^*+\beta_-^2\beta_0^*+\beta_+\beta_-^*\beta_0+\beta_+^*\beta_-\beta_0)\nonumber\\
    &\hspace{1em}-(\Gamma_+-\Gamma_-)(|\beta_+|^2-|\beta_-|^2)\beta_0+(\Gamma_++\Gamma_-)\beta_0\,,
\end{align}
where the equation of motion for $\beta_-$ can be found by applying the transformation $\mathcal{T}$ to Eq. (\ref{eq:beta_+_eom}).

\begin{figure}
    \centering
    \includegraphics{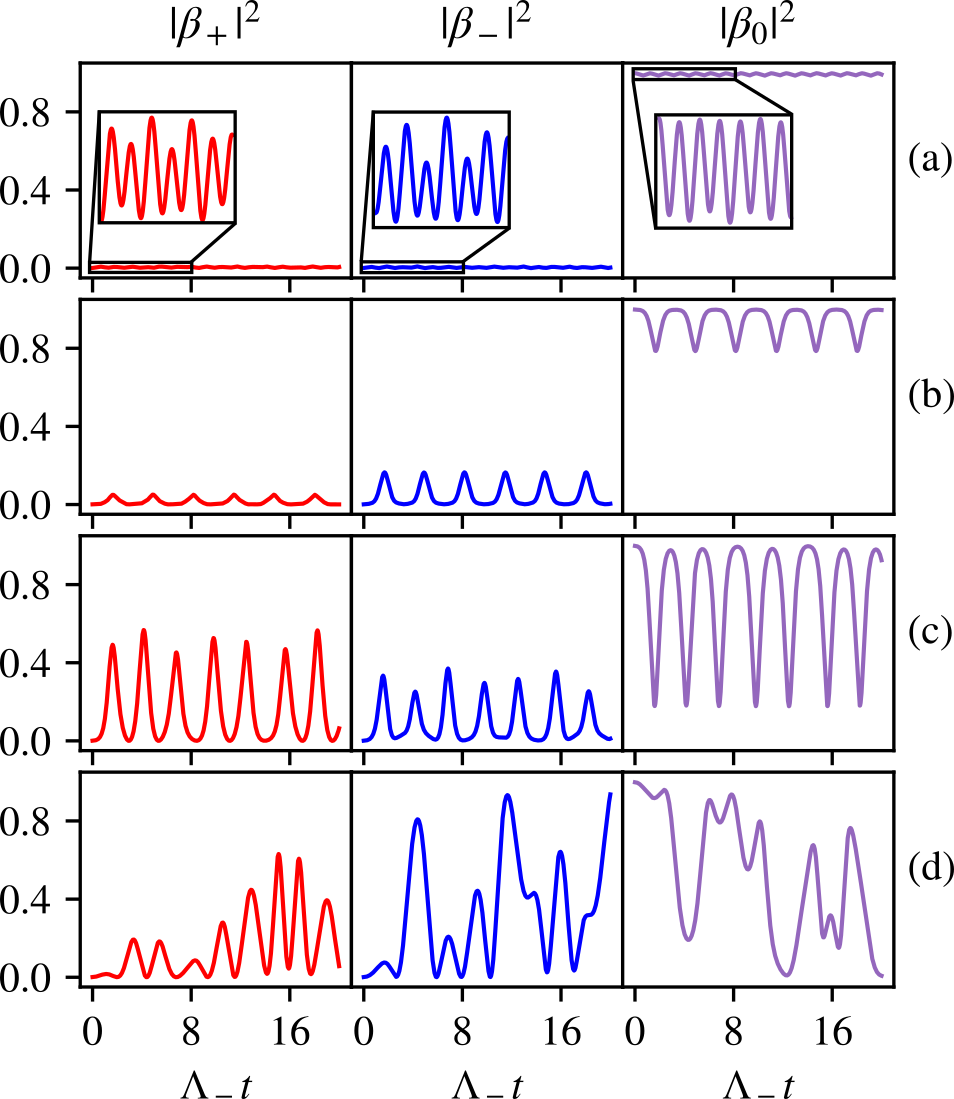}
    \caption{Numerical integration of the non-dissipative semiclassical equations of motion (\ref{eq:semiclassical_non_diss_1}) and (\ref{eq:semiclassical_non_diss_2}) for $\Lambda_+=0.5\Lambda_-$, and various $q$ and $\omega_0$ lying in different regions of parameter space. Row (a) has $q=-\Lambda_-$, $\omega_0=0$, (b) has $q=5\Lambda_-$, $\omega_0=2\Lambda_-$, (c) has $q=\Lambda_-$, $\omega_0=0$, and (d) has $q=0$, $\omega_0=\Lambda_-$. Referring to the eigenvalue {landscape in Fig. \ref{fig:open_landscapes}(a)}, row (a) lies in a region of oscillation, rows (b) and (c) lie in regions of divergence, and row (d) lies on a border between oscillation and divergence regions. Since our semiclassical treatment inherently neglects the effects of quantum fluctuations, initially unpopulated $m=\pm1$ sublevels are stationery points of the system. Thus, small seed populations are necessary to observe the dynamics. In this figure, all plots had the initial conditions $(\beta_+,\beta_-,\beta_0)=(\sqrt{0.001},\sqrt{0.001},\sqrt{0.998})$.}
    \label{fig:semiclassical_non_diss_pop_exps}
\end{figure}

However, this system of equations does not conserve the number of atoms, given by $N=|\beta_+|^2+|\beta_-|^2+|\beta_0|^2$, when $\kappa$ is nonzero. This is an artefact of the adiabatic elimination of the cavity mode, where cavity dissipation is essentially incorporated into the atomic dynamics, and thereby expressed as a changing total population. Since we already assumed and relied on number conservation, the system cannot be relied upon to give physical predictions. 
To consistently and properly include cavity dissipation we must include the cavity dynamics itself, and we do this in the following subsection. For the moment, however, we set $\kappa=0$ and the above equations reduce to
\begin{align}
    \frac{d\beta_+}{dt}&=i(\Lambda_--q-\om)\beta_++i\sqrt{\Lambda_+\Lambda_-}\beta_-\nonumber\\
    &\indent+i(\Lambda_++\Lambda_-)(\beta_+|\beta_0|^2+\beta_-^*\beta_0^2)\nonumber\\
    &\hspace{1em}+2i\sqrt{\Lambda_+\Lambda_-}(\beta_-|\beta_0|^2+\beta_+^*\beta_0^2)\,,\label{eq:semiclassical_non_diss_1}\\[0.9em]
    \frac{d\beta_0}{dt}&=i(\Lambda_++\Lambda_-)(2\beta_+\beta_-\beta_0^*+(|\beta_+|^2+|\beta_-|^2)\beta_0+\beta_0)\nonumber\\
    &\indent +2i\sqrt{\Lambda_+\Lambda_-}(\beta_+^2\beta_0^*+\beta_-^2\beta_0^*+\beta_+\beta_-^*\beta_0+\beta_+^*\beta_-\beta_0)\,.\label{eq:semiclassical_non_diss_2}
\end{align}
This system of equations can be readily integrated, and some example results are shown in Fig.~\ref{fig:semiclassical_non_diss_pop_exps} for several parameter choices. We see various types of population oscillations, ranging from roughly sinusoidal, to sharply peaked, to seemingly irregular. They can, however, be separated into two broad categories: small or large amplitude oscillations. The former involves sparse population of the $m=\pm1$ sublevels and essentially sinusoidal oscillation, while the latter involves much larger population of those sublevels, accompanied by large depletion of the $m=0$ sublevel. These large-amplitude oscillations also tend to have the same general shape -- an initial, roughly exponential increase, followed by a sharp peak, and exponential decrease.

Furthermore, the locations of these oscillations agree well with the eigenvalue landscapes of the earlier sections. The small-amplitude oscillations lie in regions of oscillation in the landscapes, while the large-amplitude oscillations lie in regions of divergence. Although here we can only confirm this result through numerical integration of the equations of motion (i.e., not through detailed analysis of its bifurcations), it is compelling and offers an explanation for the behavior of the system in regions of divergence.

\subsection{Dissipative Dynamics}\label{sec:dissipative}

Despite the insight we gained through the semiclassical description above, the issue of incorporating dissipation remains. To do so we must reintroduce the cavity mode and account for dissipation through cavity loss. If we once again take the thermodynamic limit, define the additional $c$-number
\begin{equation}
    \alpha\equiv\frac{\expv{\ah}}{\sqrt{2N}}\,,
\end{equation}
 and use the full (i.e., not adiabatically eliminated) Hamiltonian and master equation, given by Eqs.~\ref{eq:total_ham}) and (\ref{eq:full_me}) respectively, we arrive the semiclassical equations of motion
\begin{align}
    \frac{d\alpha}{dt}&=-(\kappa+i\omega)\alpha-2i\lambda_-(\beta_+\beta_0^*+\beta_-^*\beta_0)\nonumber\\
    &\hspace{1em}-2i\lambda_+(\beta_+^*\beta_0+\beta_-\beta_0^*)\label{eq:semiclassical_st}\,,\\
    \frac{d\beta_+}{dt}&=-i(q+\om)\beta_+-2i(\lambda_-\alpha+\lambda_+\alpha^*)\beta_0\,,\\
    \frac{d\beta_0}{dt}&=-2i\lambda_-(\alpha\beta_-+\alpha^*\beta_+)-2i\lambda_+(\alpha\beta_++\alpha^*\beta_-)\label{eq:semiclassical_end}\,.
\end{align}

\begin{center}
\begin{figure}
        \includegraphics{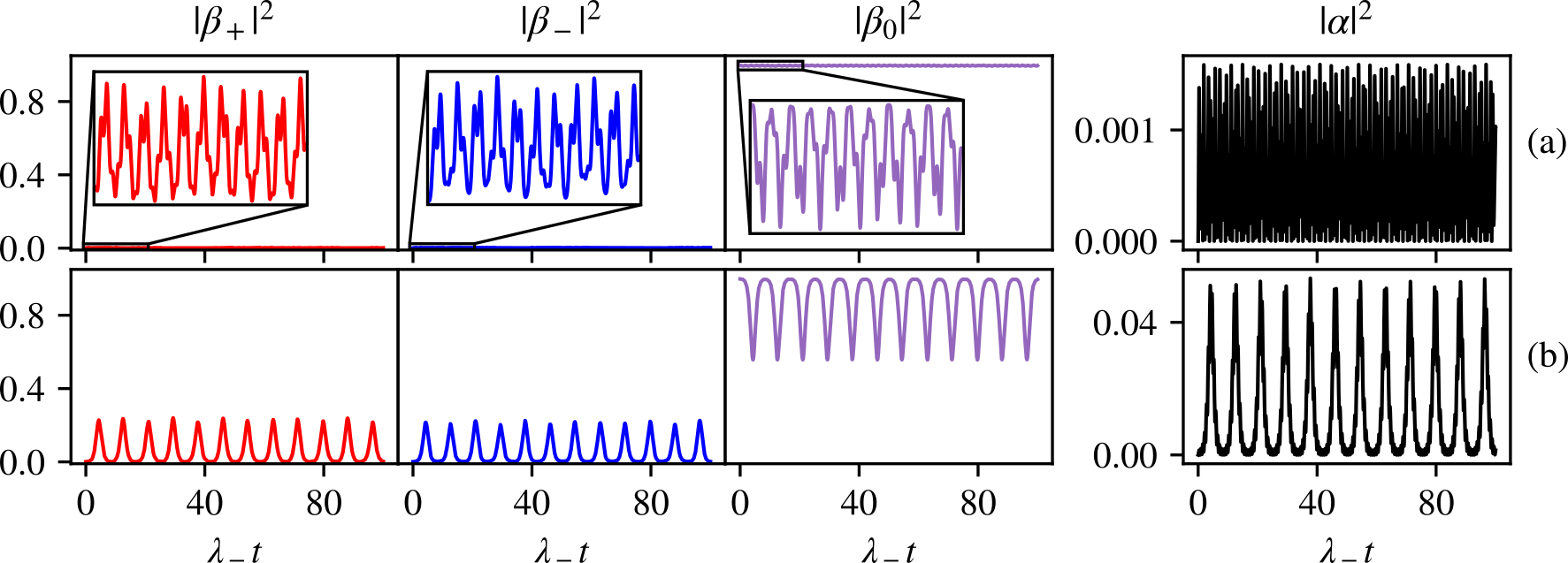}
        \caption{Atomic (left three columns) and cavity (right column) populations for $\kappa=\omega_0=0$, $\omega=10\lambda_-$, $\lambda_+=0.5\lambda_-$, and two choices of $q$: for row (a) $q$ was set to $-\lambda_-$, and for row (b) it was set to $\lambda_-$. The initial conditions for all of these plots are $(\alpha,\beta_+,\beta_-,\beta_0)=(0,\sqrt{0.001},\sqrt{0.001},\sqrt{0.998})$, corresponding to an empty cavity mode and small seed populations in the $m=\pm1$ sublevels. In relation to the eigenvalue landscapes of Section \ref{sec:closed}, row (a) corresponds to a region of oscillation, row (b) to a region of divergence.}
        \label{fig:approx_non_diss}
\end{figure}
\end{center}

In addition to conserving the number of atoms, this system allows us to study the cavity dynamics. For example, by setting $\kappa=0$ and taking $\omega$ to be large, we can approach the dispersive limit and emulate our previous semiclassical non-dissipative results. In doing so we not only confirm those results, but gain additional insight into the previous sections' regions of divergence. Figure \ref{fig:approx_non_diss} shows the cavity and atomic populations given by numerical integration of Eqs. (\ref{eq:semiclassical_st})-(\ref{eq:semiclassical_end}) under a parameter regime which approximates the dispersive limit. The atomic populations once again follow either small or large amplitude oscillation, in agreement with the eigenvalue landscapes. Furthermore, as one might expect, at every significant depletion of the $m=0$ sublevel during large-amplitude oscillations a burst of photons is generated in the cavity. 

\begin{center}
\begin{figure}[t]
        \centering
        \includegraphics{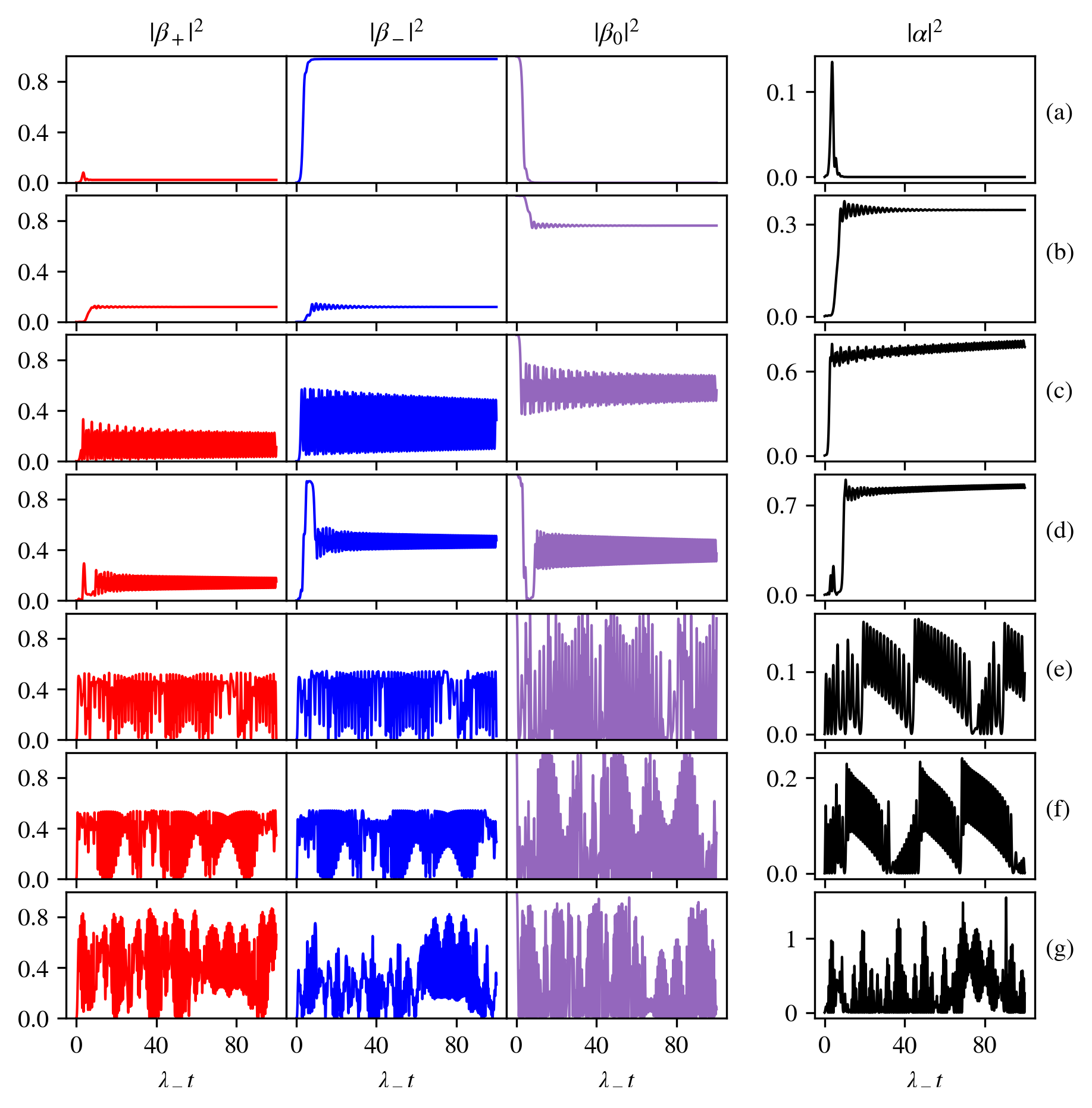}
        \caption{Atomic (three left columns) and cavity (right column)  populations for a variety of  parameter sets. Plots in rows (a-d) have $\kappa=\omega=2\lambda_-$, with (a) corresponding to $q=\lambda_-$, $\omega_0=\lambda_+=0$, (b) to $q=2\lambda_-$, $\omega_0=0$, $\lambda_+=0.5\lambda_-$, (c) to $q=2\lambda_-$, $\omega_0=\lambda_+=\lambda_-$, and (d) to $q=-\lambda_-$, $\omega_0=\lambda_+=\lambda_-$. Rows (e-g) have $\omega_0=\kappa=\omega$ and $\lambda_+=2\lambda_-$ and varying values of $q$ and $\lambda_-$: (e) has $q=0$, $\lambda_-=2\omega$, (f) has $q=0$, $\lambda_-=3\omega$, and (g) has $q=0.5$, $\lambda_-=2\omega$ The initial conditions for all of these plots are as in Fig. \ref{fig:approx_non_diss}, $(\alpha,\beta_+,\beta_-,\beta_0)=(0,\sqrt{0.001},\sqrt{0.001},\sqrt{0.998})$. Rows (a-c) demonstrate the three known phases of the Dicke mode: the normal phase (a), superradiance (b), and oscillatory superradiance (c). Plot (d) additionally shows multistability of the normal and oscillatory phases, where the system initially behaves as in the normal phase, but eventually settles into a limit cycle of the oscillatory phase. Rows (e-g) show three chaotic dynamics. Note that rows (e) and (f) show similar cavity populations to those in Ref. \cite{stitely}, while row (g) are markedly different.}
        \label{fig:dicke_phases}
\end{figure}
\end{center}

Moreover, we now have the freedom to explore dissipative scenarios. Figs.~\ref{fig:dicke_phases}(a-d) show the system evolution for four different parameter sets that correspond to the known phases of the Dicke model. In the normal phase, atoms may undergo a one-way transition from the $m=0$ sublevel to the other sublevels, accompanied by a burst of photons and subsequent decay towards a steady state with no photons in the cavity (Fig.~\ref{fig:dicke_phases}(a)). In the superradiant phase, both the cavity and atomic modes settle into a steady state with constant populations, corresponding to balanced atomic transitions (Fig.~\ref{fig:dicke_phases}(b)). In oscillatory superradiance the steady state is akin to the superradiant phase, but involves oscillating atomic and cavity populations (Fig.~\ref{fig:dicke_phases}(c)). Lastly, we see signatures of normal-superradiant bistability, where the system appears to pass close to the normal phase equilibrium, but eventually reaches the superradiant equilibrium (Fig.~\ref{fig:dicke_phases}(d)).

These are all in agreement with previous analyses of the two-level-atom version of the Dicke model \cite{stitely}, which not only mapped these phases, but also regions of chaotic behaviour. It therefore comes as no surprise that we also observe chaos, which specifically occurs when the counter-rotating terms in the Hamiltonian dominate, i.e., when $\lambda_+>\lambda_-$. Figs.~\ref{fig:dicke_phases}(e-g) show variations of the chaos we observe: two are very similar to the two-level system chaos, where we see sudden spikes and slower decay of the cavity population. However, the other, with $q\neq 0$, is quite different; such spikes and decays are not as easily visible, and the cavity population appears to fluctuate randomly.

\onecolumngrid
\twocolumngrid

\onecolumngrid
\twocolumngrid

These chaotic behaviors are also manifest in the atomic dynamics. Although population time-series give a useful context to some behaviors, they become less useful when the dynamics are chaotic, given they only provide a one-dimensional projection of our eight-dimensional dynamical system. A more useful representation of the atomic dynamics is a plot of the system trajectory in the space spanned by the spin expectations (hence dubbed the ``spin-space'') $s_x=\expv{\Sh_x}$, $s_y=\expv{\Sh_y}$, and $s_z=\expv{\Sh_z}$. Since such a representation is a three dimensional projection of the full system, its trajectories are allowed to self-intersect (though they do not self-intersect in the full eight-dimensional phase space), but they nevertheless provide information on attractors and equilibria in the system, and have indeed been employed in investigations of the two-level system \cite{stitely}.

\begin{figure}
    \centering
    \includegraphics{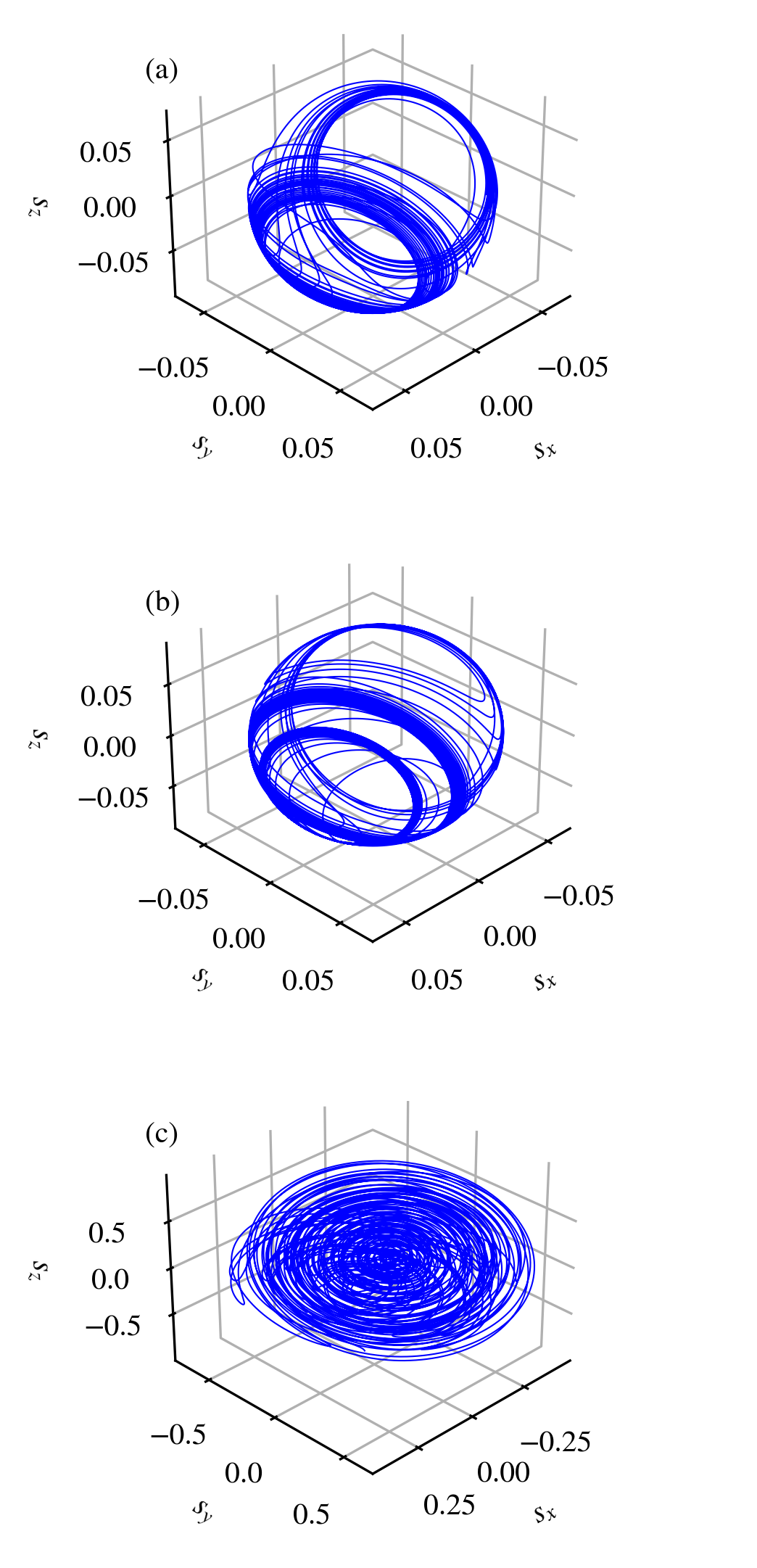}
    \caption{Three types of chaotic spin-space trajectories: two-attractor in (a), three-attractor in (b), and non-Bloch sphere in (c). The parameters for (a-c) are the same as those for Figs.~\ref{fig:dicke_phases}(e-g), respectively. The initial conditions for all of these are $(\alpha,\beta_+,\beta_-,\beta_0)=(0,\sqrt{0.001},\sqrt{0.001},\sqrt{0.998})$, though the trajectory is only plotted from $\lambda_-t=100$ to $\lambda_-t=200$ for clarity.}
    \label{fig:spin_space_chaos}
\end{figure}

In Fig.~\ref{fig:spin_space_chaos}, we plot the spin-space trajectories of the chaotic dynamics shown in Fig.~\ref{fig:dicke_phases}, which characterize three forms of chaos in our system. First, in Fig.~\ref{fig:spin_space_chaos}(a), we see a trajectory typical of the two-level system: a symmetric (about reflections along the $s_x=s_y$ diagonal) chaotic attractor is present and the system maintains a constant spin-length (i.e., the trajectory lies on the Bloch sphere). In Fig.~\ref{fig:spin_space_chaos}(b), we see an asymmetric chaotic attractor that also follows the Bloch sphere, which most likely arises due to the additional atomic degree of freedom compared with the two-level model. Lastly, in Fig.~\ref{fig:spin_space_chaos}(c) the attractor is not confined to the Bloch sphere.

{This may be explained by considering the total ensemble spin, which is only conserved when atomic transitions to the $m=+1$ magnetic sublevel result in energy changes equal in magnitude but opposite in sign to transitions to the $m=-1$ sublevel. In other words, the spin-length is only conserved when $q=0$. One can also see this by noting that $\hat{S}^2=\hat{S}^2_z+\hat{S}_z+\hat{S}_-\hat{S}_+$ commutes with all of the Hamiltonian terms except the term pertaining to the quadratic Zeeman shift. More specifically, the operator $\hat{b}_+^\dag\hat{b}_++\hat{b}_-^\dag\hat{b}_-$ commutes with $\hat{S}_z=\hat{b}_+^\dag\hat{b}_+-\hat{b}_-^\dag\hat{b}_-$, but does not commute with $\hat{S}_-\hat{S}_+\propto(\hat{b}_0^\dag\hat{b}_++\hat{b}_-^\dag\hat{b}_0)(\hat{b}_+^\dag\hat{b}_0+\hat{b}_0^\dag\hat{b}_-)$.} Therefore, the two trajectories in Figs.~\ref{fig:spin_space_chaos}(a,b) follow the Bloch sphere, while the trajectory in Fig.~\ref{fig:spin_space_chaos}(c) does not.

In order to construct a preliminary map of the various phases and chaotic behaviour in parameter space, we use a simple numerical scheme to analyse the mean and any nonzero, steady-state oscillation amplitude. At each point in a grid of $\lambda_+$ and $\lambda_-$ values, we numerically integrate the system of equations, extract the cavity population over a late portion of the trajectory, and calculate the mean and oscillation amplitude (if any) of the sample. The mean is calculated by simply averaging over the sample, and the amplitude calculated by taking the difference between the maximum and minimum values and dividing it by two. It should be noted that the mean and oscillattion amplitude of the steady-state cavity population are sufficient to characterize the phase of the system: in the normal phase both the mean and amplitude will be zero, in the superradiant phase the mean will be nonzero but the oscillation amplitude is zero, and in the oscillatory phase both will be nonzero. Since no steady-state is reached in chaotic behaviour, we expect it to display seemingly random fluctuations in both the mean and amplitude.

We must note, however, that these characterizations only hold true under idealized conditions, and hinge on the system reaching the steady-state (when one exists). Therefore, if a particular trajectory takes too long to reach steady-state, or if the portion of the trajectory over which we perform our average is too short, the calculated mean and amplitude may not faithfully represent the true phase. In addition, since the calculation evolves the system in discrete timesteps, if the timestep is exactly a half-integer multiple of an oscillation period, the amplitude may register incorrectly as zero. Lastly, the calculation is somewhat reliant on the initial conditions, so the emergence of certain phases in a regime of multistability may not be detected. Nevertheless, these caveats will not influence the entirety of our calculations, so we can rely on them to give us at least a rough map for the location of each phase.

\begin{figure}
    \centering
    \includegraphics{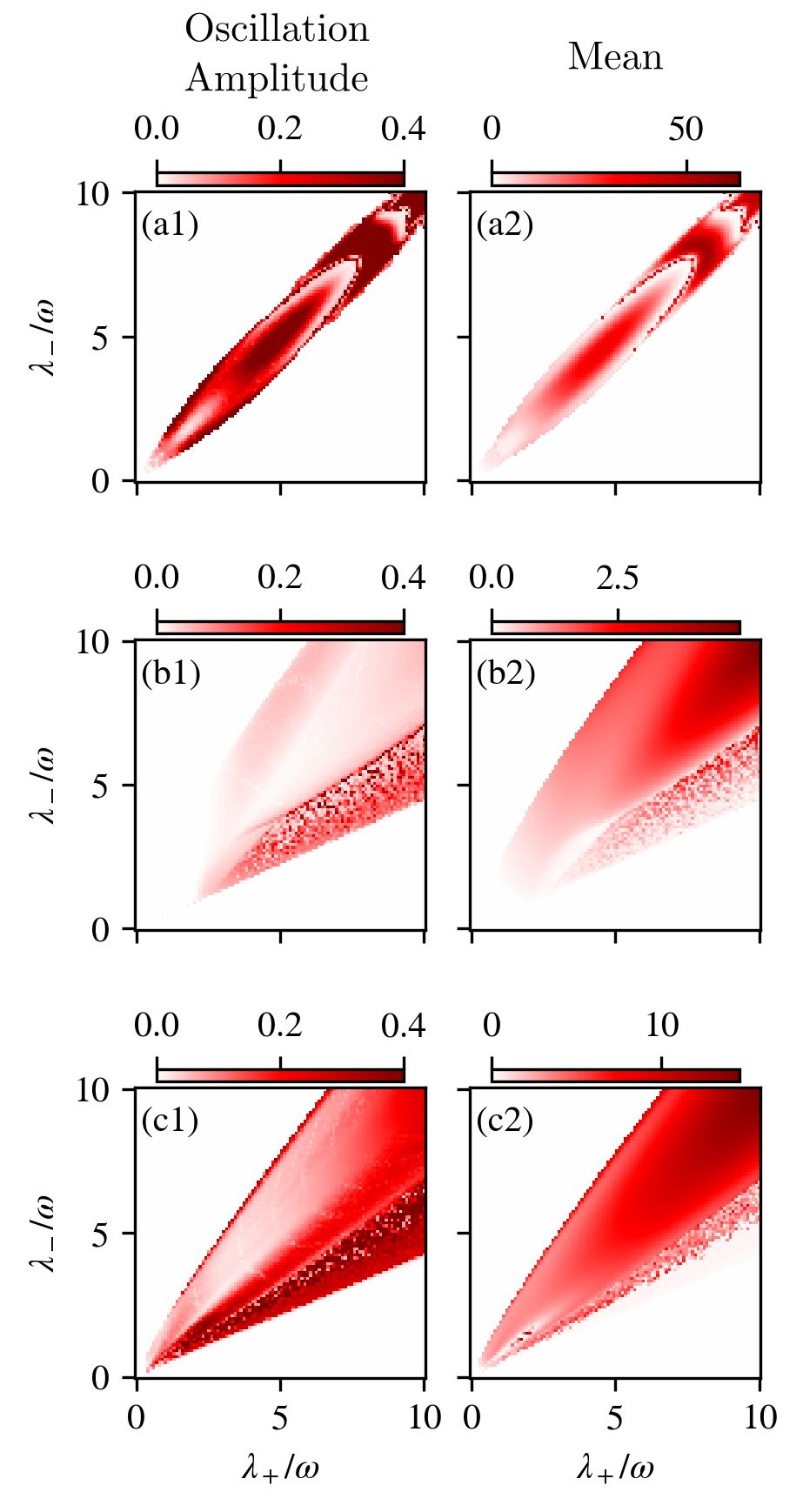}
    \caption{Oscillation amplitudes (a-c1) and means (a-c2) of long-time trajectory portions (from $\omega t=150$ to $\omega t=200$) for a variety of $q$ and $\omega_0$ values, pertaining to different energy level structures. Plots (a1,2) correspond to $q=-\omega$, $\omega_0=0$, which is a symmetric energy level structure, but has an initially inverted population given $q$ is negative.  Plots (b1,2) correspond to $q=0$, $\omega_0=\omega$, which is asymmetrical but follows a similar phase diagram to the two-level model, given $q=0$. Plots (c1,2) correspond to $q=\omega_0=0.5\omega$, where one sublevel ($m=-1$ in this case) has the same energy as the $m=0$ sublevel, while the other ($m=1$) has a higher energy. Initial conditions for all calculations are  $(\alpha,\beta_+,\beta_-,\beta_0)=(0,\sqrt{0.001},\sqrt{0.001},\sqrt{0.998})$.}
    \label{fig:phase_landscapes}
\end{figure}

Fig.~\ref{fig:phase_landscapes} shows the results of three such calculations, for choices of $q$ and $\omega_0$ pertaining to different energy-level structures. Regardless of the specific values of $q$ and $\omega_0$, we see regions where each phase is displayed, as per the criteria above, with the most notable separation between the normal phase and superradiance. In Figs.~\ref{fig:phase_landscapes}(b1,b2) and \ref{fig:phase_landscapes}(c1,c2) there is also a seemingly sharp boundary between chaos and the other phases; in plots \ref{fig:phase_landscapes}(b1,2) specifically, where $q=0$, the boundary appears to agree well with the equivalent phase diagram for the two-level model \cite{stitely}. Lastly, we note that plots \ref{fig:phase_landscapes}(a1,2) show no chaotic regions, possibly owing to the symmetrical energy level structure (i.e., $\omega_0=0$) being considered.

%: --------------------------------------------------------------
%:                  Conclusion
% --------------------------------------------------------------
\section{Conclusion and Outlook}\label{sec:conclusion}

In this initial work we have investigated the spin-1 version of the Dicke model in a generalized setting, under both quantum-mechanical and semiclassical lenses, by introducing to it a quadratic Zeeman shift as a tunable parameter. By varying this parameter and the various other parameters of the Dicke model, we were able to explore arbitrary atomic energy-level configurations and unbalanced rotating and counter-rotating atom-cavity coupling terms in dissipative scenarios, and thereby extend the model beyond its traditional two-level flavour.

We began by making the undepleted $m=0$ mode approximation in the closed model, with the cavity mode adiabatically eliminated, and analysing the resultant operator equations of motion. These equations were all linear and coupled, allowing us to analyse their eigenvalues in order to determine the stability of their solutions. This analysis revealed regions in parameter space where solutions oscillated or diverged, and whose boundaries we found analytically.

We then introduced dissipation to the model through cavity loss, and used a master equation approach to arrive at a set of equations of motion for various operator moments. We performed an eigenvalue analysis on the first order moments as before, and discovered the inclusion of dissipation caused widespread divergence throughout parameter space. Numerical integration of the second-order moment equations revealed populations increased exponentially, though in the aforementioned regions of oscillation this increase was slower.

Given atomic populations are not permitted to increase indefinitely in realistic scenarios, in order to proceed further we needed to relax the undepleted $m=0$ mode approximation and take the thermodynamic limit. Doing so allowed us to factorize expectations and derive nonlinear equations of motion for first-order moments. These only held in the non-dissipative case, and showed that in the aforementioned regions of divergence, the atomic populations oscillated with large amplitudes, and the $m=0$ sublevel became significantly depleted.

In incorporating dissipation into this semiclassical description, we needed to relax the adiabatic elimination of the cavity mode, and consider the full Hamiltonian and master equation. We once again arrived at nonlinear semiclassical equations of motion, which allowed us to not only investigate dissipative atomic dynamics in the regions of divergence, but study their accompanying cavity dynamics. Using a parameter regime which emulated the previous, adabatically-eliminated, non-dissipative system, we found that with each significant depletion of the $m=0$ sublevel, a burst of photons was generated in the cavity.

Our full semiclassical model additionally displayed all three of the standard, known Dicke model phases, as well as multistability and chaos, similar to the two-level model. Our model, however, showed forms of chaos not seen in the two-level model, namely asymmetric trajectories, and trajectories which do not conserve the total spin-length. We created rough maps of the different phases in varying parameter regimes, where, notably, we see sharp boundaries between the normal phase, superradiance, and chaos.

The richness and variety of behaviors observed in our semiclassical analysis show that more research needs to be done in order to fully characterize the system. For example, bifurcation theory could be used to determine when the normal-superradiant transition occurs analytically, and numerical methods, such as continuation, could be used to map the locations of superradiant-oscillatory transitions, the emergence of chaos, and types of chaotic behaviour \cite{stitely}. This remains the approach of ongoing work \cite{Adiv2024}.

Lastly, the emergence of these semiclassical behaviors, and chaos in particular, from the quantum mechanical description of the model can be further investigated. Quantum fluctuations in this transition have led to interesting dynamics in the two-level system \cite{Stitely2020}, and with the additional atomic degree of freedom in our model they may give rise to further interesting and novel behavior.

\begin{acknowledgments}
This research was supported in part by grants NSF PHY-1748958 and PHY-2309135 to the Kavli Institute for Theoretical Physics (KITP). 
\end{acknowledgments}

%: --------------------------------------------------------------
%:                  Appendices
% --------------------------------------------------------------
\onecolumngrid

\appendix
\section{Moment Equations for Second-Order Operators in the Open System}\label{appdx:mom_eqns}

As outlined in Sections \ref{sec:closed} and \ref{sec:pop_dyn}, we only need to find four second-order moment equations to find the remaining six. We chose to find these for $\bh_+^\dag\bh_+$, $\bh_+^2$, $\bh_+\bh_-$, and $\bh_+\bh_-^\dag$, which are, respectively,
\begin{align}
    \frac{d\expv{\bh_+^\dag\bh_+}}{dt}&=i(\Lambda_++\Lambda_-)(\expv{\bh_+^\dag\bh_-^\dag}-\expv{\bh_+\bh_-})+2i\sqrt{\Lambda_+\Lambda_-}(\expv{\bh_+^{\dag2}}-\expv{\bh_+^2}+\expv{\bh_+^\dag\bh_-}-\expv{\bh_+\bh_-^\dag})\nonumber\\
    &\hspace{1em}+(\Gamma_+-\Gamma_-)(2\expv{\bh_+^\dag\bh_+}+\expv{\bh_+\bh_-}+\expv{\bh_+^\dag\bh_-^\dag})+2\Gamma_+\,,\label{eq:open_2nd_order_eom_st}\\\nonumber\\
    \frac{d\expv{\bh_+^2}}{dt}&=-2i(q+\om)\expv{\bh_+^2}+2(\Gamma_+-\Gamma_-+i(\Lambda_++\Lambda_-))(\expv{\bh_+^2}+\expv{\bh_+\bh_-^\dag})\nonumber\\
    &\hspace{1em}+4i\sqrt{\Lambda_+\Lambda_-}(\expv{\bh_+^\dag\bh_+}+\expv{\bh_+\bh_-})+2i\sqrt{\Lambda_+\Lambda_-}-2\sqrt{\Gamma_+\Gamma_-}\,,\\\nonumber\\
    \frac{d\expv{\bh_+\bh_-}}{dt}&=-2i(q-\Lambda_+-\Lambda_-)\expv{\bh_+\bh_-}+i(\Lambda_++\Lambda_-)(\expv{\bh_+^\dag\bh_+}+\expv{\bh_-^\dag\bh_-})-(\Gamma_+-\Gamma_-)(\expv{\bh_+^\dag\bh_+}-\expv{\bh_-^\dag\bh_-})\nonumber\\
    &\hspace{1em}+2i\sqrt{\Lambda_+\Lambda_-}(\expv{\bh_+^2}+\expv{\bh_-^2}+\expv{\bh_+\bh_-^\dag}+\expv{\bh_+^\dag\bh_-})-(\Gamma_++\Gamma_-)+i(\Lambda_++\Lambda_-)\,,\\\nonumber\\
    \frac{d\expv{\bh_+\bh_-^\dag}}{dt}&=-2i\om\expv{\bh_+\bh_-^\dag}+(\Gamma_+-\Gamma_-+i(\Lambda_++\Lambda_-))(\expv{\bh_-^{\dag2}}-\expv{\bh_+^2})\nonumber\\
    &\hspace{1em}+2i\sqrt{\Lambda_+\Lambda_-}(\expv{\bh_-^\dag\bh_-}-\expv{\bh_+^\dag\bh_+}+\expv{\bh_+^\dag\bh_-^\dag}-\expv{\bh_+\bh_-})+2\sqrt{\Gamma_+\Gamma_-}\,.
\end{align}
The remaining equations, found either by conjugation or using the transformation $\mathcal{T}$, are
\begin{align}
    \frac{d\expv{\bh_-^\dag\bh_-}}{dt}&=i(\Lambda_++\Lambda_-)(\expv{\bh_+^\dag\bh_-^\dag}-\expv{\bh_+\bh_-})+2i\sqrt{\Lambda_+\Lambda_-}(\expv{\bh_-^{\dag2}}-\expv{\bh_-^2}+\expv{\bh_+\bh_-^\dag}-\expv{\bh_+^\dag\bh_-})\nonumber\\
    &\hspace{1em}+(\Gamma_--\Gamma_+)(2\expv{\bh_-^\dag\bh_-}+\expv{\bh_+\bh_-}+\expv{\bh_+^\dag\bh_-^\dag})+2\Gamma_-\,,\\\nonumber\\
    \frac{d\expv{\bh_-^2}}{dt}&=-2i(q-\om)\expv{\bh_-^2}+2(\Gamma_--\Gamma_++i(\Lambda_++\Lambda_-))(\expv{\bh_-^2}+\expv{\bh_+^\dag\bh_-})\nonumber\\
    &\hspace{1em}+4i\sqrt{\Lambda_+\Lambda_-}(\expv{\bh_-^\dag\bh_-}+\expv{\bh_+\bh_-})+2i\sqrt{\Lambda_+\Lambda_-}-2\sqrt{\Gamma_+\Gamma_-}\,,\\\nonumber\\
    \frac{d\expv{\bh_+^{\dag2}}}{dt}&=2i(q+\om)\expv{\bh_+^{\dag2}}+2(\Gamma_+-\Gamma_--i(\Lambda_++\Lambda_-))(\expv{\bh_+^{\dag2}}+\expv{\bh_+^\dag\bh_-})\nonumber\\
    &\hspace{1em}-4i\sqrt{\Lambda_+\Lambda_-}(\expv{\bh_+^\dag\bh_+}+\expv{\bh_+^\dag\bh_-^\dag})-2i\sqrt{\Lambda_+\Lambda_-}-2\sqrt{\Gamma_+\Gamma_-}\,,\\\nonumber\\
    \frac{d\expv{\bh_-^{\dag2}}}{dt}&=2i(q-\om)\expv{\bh_-^{\dag2}}+2(\Gamma_--\Gamma_+-i(\Lambda_++\Lambda_-))(\expv{\bh_-^{\dag2}}+\expv{\bh_+\bh_-^\dag})\nonumber\\
    &\hspace{1em}-4i\sqrt{\Lambda_+\Lambda_-}(\expv{\bh_-^\dag\bh_-}+\expv{\bh_+^\dag\bh_-^\dag})-2i\sqrt{\Lambda_+\Lambda_-}-2\sqrt{\Gamma_+\Gamma_-}\,,\\
    \frac{d\expv{\bh_+^\dag\bh_-^\dag}}{dt}&=2i(q-\Lambda_+-\Lambda_-)\expv{\bh_+^\dag\bh_-^\dag}-i(\Lambda_++\Lambda_-)(\expv{\bh_+^\dag\bh_+}+\expv{\bh_-^\dag\bh_-})-(\Gamma_+-\Gamma_-)(\expv{\bh_+^\dag\bh_+}-\expv{\bh_-^\dag\bh_-})\nonumber\\
    &\hspace{1em}-2i\sqrt{\Lambda_+\Lambda_-}(\expv{\bh_+^{\dag2}}+\expv{\bh_-^{\dag2}}+\expv{\bh_+^\dag\bh_-}+\expv{\bh_+\bh_-^\dag})-(\Gamma_++\Gamma_-)-i(\Lambda_++\Lambda_-)\,,\\\nonumber\\
    \frac{d\expv{\bh_+^\dag\bh_-}}{dt}&=2i\om\expv{\bh_+^\dag\bh_-}+(\Gamma_+-\Gamma_--i(\Lambda_++\Lambda_-))(\expv{\bh_-^2}-\expv{\bh_+^{\dag2}})\nonumber\\
    &\hspace{1em}-2i\sqrt{\Lambda_+\Lambda_-}(\expv{\bh_-^\dag\bh_-}-\expv{\bh_+^\dag\bh_+}+\expv{\bh_+\bh_-}-\expv{\bh_+^\dag\bh_-^\dag})+2\sqrt{\Gamma_+\Gamma_-}\,.\label{eq:open_2nd_order_eom_fin}
\end{align}
\twocolumngrid
%: --------------------------------------------------------------
%:                  References
% --------------------------------------------------------------
\bibliography{refs.bib}\label{sec:refs}

\end{document}